\documentclass[]{pasj01}
\draft
\usepackage{lscape}
\usepackage{lineno}

\Received{}
\Accepted{}
 
\begin{document} 

\title{ 
Relationship between Gamma-ray loudness and \\X-ray spectra of Radio Galaxies}

\author{Taishu \textsc{Kayanoki}\altaffilmark{1,*}}
\altaffiltext{1}{Department of Physics, Graduate School of Advanced Science and Engineering, Hiroshima University 1-3-1 Kagamiyama, Higashi-Hiroshima, Hiroshima 739-8526, Japan}
\email{kayanoki@astro.hiroshima-u.ac.jp}

\author{Yasushi \textsc{Fukazawa}\altaffilmark{1}}



\maketitle

\begin{abstract}
The {\it Fermi} satellite has detected $\sim$60 radio galaxies (RGs). In this study, we investigate the difference in the properties of X-ray spectra between GeV-loud RGs and GeV-quiet RGs. Our sample comprises 68 objects: 36 RGs detected with {\it Fermi} and 32 RGs not detected with gamma rays. We analyzed the X-ray spectra of these 68 objects using data from the {\it XMM-Newton}, {\it Chandra}, {\it NuSTAR}, and {\it Swift} satellites. Our results show that most GeV-loud RGs do not exhibit significant absorption, while $\sim$50\% of the GeV-quiet RGs exhibit significant absorption. This suggests that the jet of GeV-loud RGs is viewed from a small angle, and thus the emission is not easily blocked by the torus. Moreover, we reported that RGs with a heavy absorption are mostly in the X-ray luminosity range of $10^{43}-10^{45}$erg s$^{-1}$; however, few RGs with lower and higher luminosity suffer from heavy absorption. This is the same trend as that of Seyfert galaxies.
\end{abstract}

\section{Introduction}
Radio galaxies (RGs) are radio-loud Active galactic nuclei (AGN) with jets misaligned to the line of sight. Therefore, unlike blazars in which beamed jet emissions are dominant, RGs are objects in which emission from both disk/corona and jet is observed. RGs are classified in FR-I and FR-II based on their radio luminosity and morphology (\citet{Fanaroff}). Radio luminosity of FR-Is is less than $10^{26}~\rm{W\,Hz^{-1}}$ at 178 MHz, and radio flux is higher near the core and fades toward the outer region. Radio luminosity of FR-IIs is higher than $10^{26}~\rm{W\,Hz^{-1}}$ at 178 MHz, and radio flux is low close to the core and becomes bright toward the lobe edge region with bright hot spots. FR-I is considered to have a low mass accretion rate, while FR-II is considered to have a high mass accretion rate. Others that do not fall into this classification are Compact Steep Spectrum (CSS). CSS is a radio source associated with an AGN, which is characterized by its small radio emission size and steep slope of the radio spectrum.

Because jets of RGs are viewed at a larger angle from the line of sight, the gamma-ray luminosity of RGs is lower than that of blazars by several orders of magnitude, and the number of detections in the GeV band is small. Therefore, it was difficult to statistically study the difference in properties between GeV-loud RGs and GeV-quiet RGs. However, with the 4th Fermi-LAT catalog (4FGL-DR2) \citep{Ballet}, the number of RGs detected in the GeV gamma-ray band detections has increased and statistical analysis can be performed. 
Hereafter, we call {\it Fermi}-detected RGs as GeV-loud RGs and other RGs as GeV-quiet RGs.
However, the number of RGs detected with {\it Fermi} is around 10\% compared with those detected in the radio band, based on the comparison of logN-logS relation and flux ratio between radio and GeV gamma-ray bands \citep{Inoue11}.
As shown later in our sample RGs, radio-bright RGs are not necessarily detected in the gamma-ray band; only $\sim$10\% of the RGs seem to be GeV-loud.
This could be caused by the difference in viewing angle of jets; for GeV-loud RGs, their jet is viewed with a smaller viewing angle.
\citet{Inoue11} pointed out this possibility.
\citet{Fukazawa} systematically compiled the angle of the jets to be 10--30 degrees by the multi-wavelength Spectral Energy Distribution (SED) modeling with the Synchrotron Self-Compton (SSC) model for 10 RGs listed in the first Fermi-LAT catalog \citet{1FGL}. However, their uncertainties are large.
In addition, SED of RGs consists of not only jet emission component but also other components from the host galaxy, disk/corona, and interstellar medium and thus SED modeling is often difficult.
In some cases, a viewing angle of jets was estimated from the radio flux ratio between two sides of the jet \citep{Fujita17}, but such cases of estimation are still limited.

AGN emission from the central region close to the supermassive black hole (SMBH) can be observed as a power-law shape continuum emission in the X-ray band. AGN central region is often surrounded by the torus. When the central region viewed through this matter, X-ray continuum emission is observed as an absorbed power-law shape. About 80\% of Seyfert galaxies with X-ray luminosity of $\sim10^{43}\rm{erg\,s^{-1}}$ show a large absorption column density of  $N\rm{_H}>10^{22}\rm{cm^{-2}}$; however, less than 50\% of Seyfert galaxies with X-ray luminosity of $<10^{42}\rm{erg\,s^{-1}}$ or $>10^{44}\rm{erg\,s^{-1}}$ do \citep{Beckmann,Burlon}.

Since the jet direction is considered to be almost the same as the torus axis, we can know about the viewing angle of jets by using torus X-ray absorption.
In previous reports (e.g. \cite{Kang,Panessa}), the nature of the X-ray spectra of RGs has been statistically investigated, and a similar trend was found for RGs compared with radio-quiet AGNs. 
However, no comparison has been made between GeV-loud and GeV-quiet RGs, since the number of GeV-loud RGs has been limited.
Therefore, a large number of RGs in the 4FGL-DR2 catalog for the first time enables us to compare the X-ray absorption between GeV-loud and GeV-quiet RGs.
In this paper, we systematically analyzed available X-ray observational data of RGs and studied the dependence of X-ray absorption on gamma-ray detection.
Cosmological parameters ($H_0$,\,$\Omega_m$,\,$\Omega_\Lambda$) used in this study are $H_0=70\,\rm{km\,s^{-1}Mpc^{-1}}$, $\Omega_M=0.3$, $\Omega_{\Lambda}=0.7$.

\section{Sample and X-ray data}
\subsection{Sample Radio Galaxies}
\begin{figure}[ht]
  \begin{minipage}[h]{0.5\hsize}
    \centering
    \includegraphics[height=7.0cm,width=8cm]{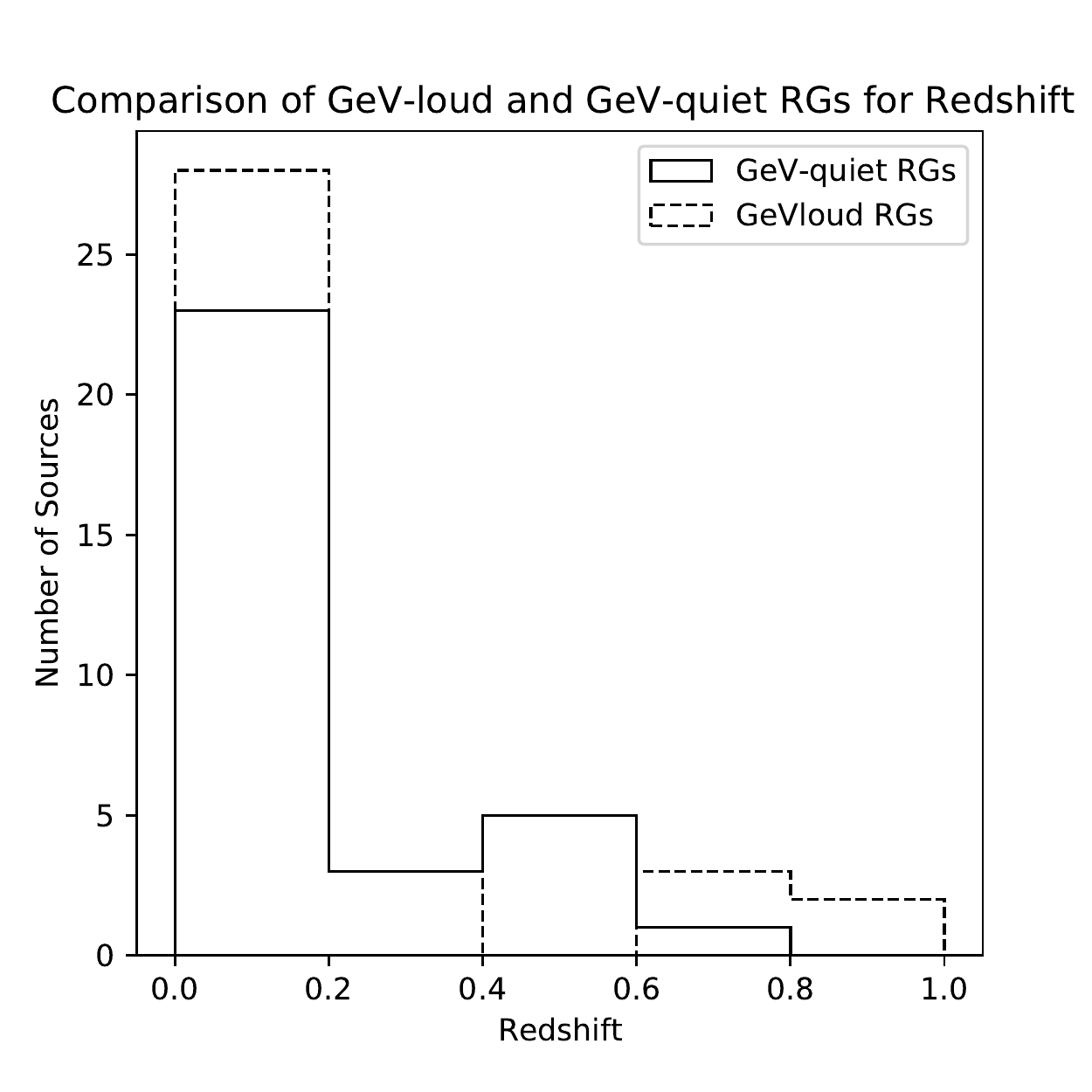}
    \end{minipage} 
  \begin{minipage}[h]{0.5\hsize}
    \centering
    \includegraphics[height=7.0cm,width=8cm]{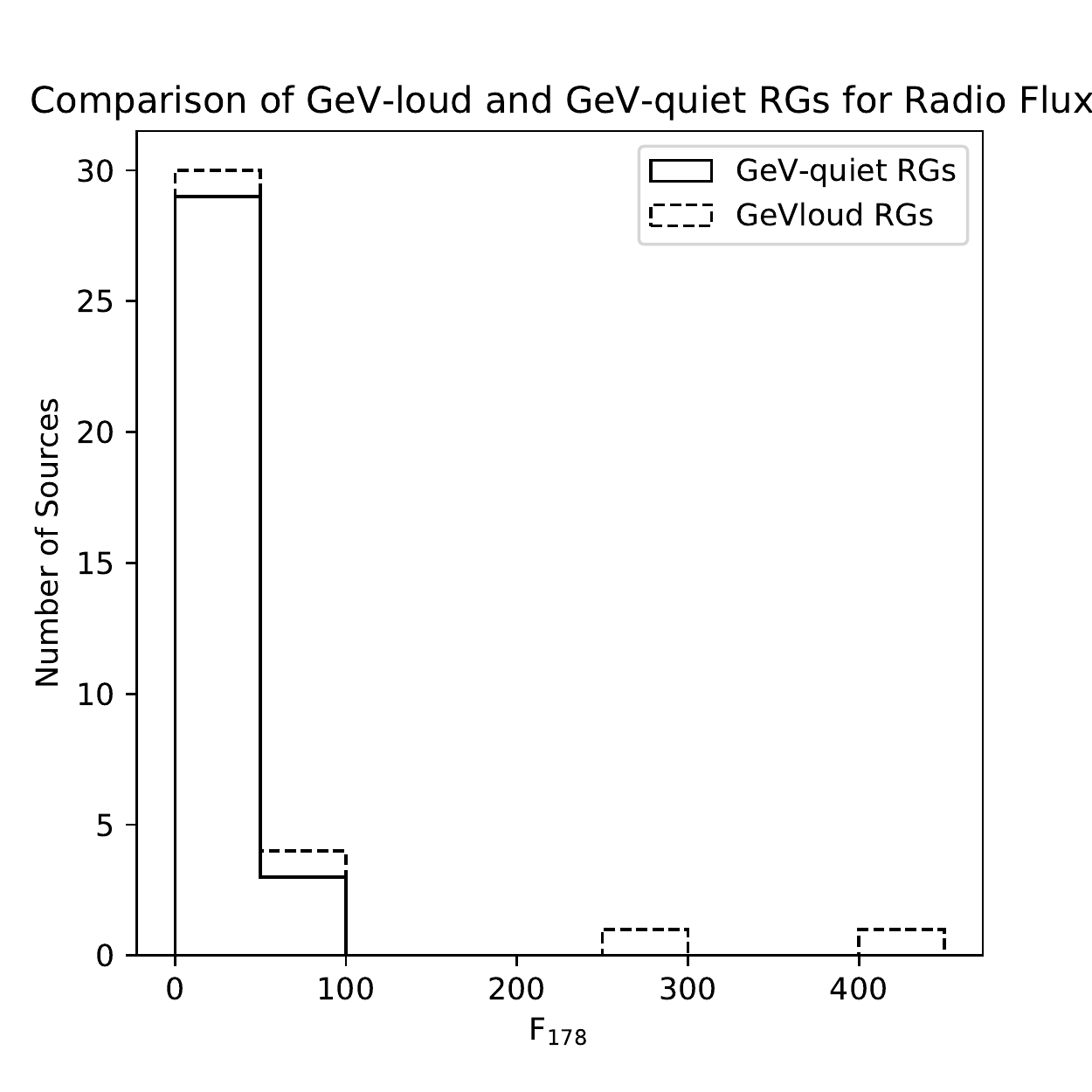}
  \end{minipage} 
 
\caption{Distributions of redshift (left) and radio flux at 178 MHz (right) of GeV-loud RGs and GeV-quiet RGs in our sample. The solid line shows GeV-quiet RGs and the dashed line GeV-loud RGs. }
 \label{hist}
\end{figure}

We use the 4FGL-DR2 catalog to select GeV-loud RGs, and radio flux-limited sample RG \citep{Mingo,Massaro} to select GeV-quiet RGs. The 4FGL-DR2 catalog was published \citep{Ballet} as the 4th Fermi/LAT Gamma-ray source catalog. The 4FGL-DR2 catalog contains 61 misaligned AGNs, of which 43 are classified as RG, 5 as CSS, and 11 as AGN\footnote{One of object classes name in 4FGL-DR2. This class corresponds to non-blazar AGNs whose existing data do not allow an unambiguous determination of their AGN types.}. Also, \citet{Mingo} listed 45 RGs with fluxes of $>$2 Jy at 2.7 GHz. These are southern RGs and have a flat radio spectrum. \citet{Massaro} compiled 93 RGs in the 3CR catalogue that have been observed in {\it Chandra}. We selected RGs for which X-ray data of the {\it XMM-Newton}, {\it Chandra}, {\it Swift}, or {\it NuSTAR} are available. If the X-ray emission can not be clearly seen at the RG position or most of X-ray emission is extended, such RGs were excluded from our sample.
We defined GeV-loud RGs in our sample as those listed in the 4FGL-DR2 catalog and selected 38 GeV-loud RGs. For GeV-quiet RGs, we first selected RGs from \citet{Mingo}, but their sample is poor in FR-Is while GeV-loud RGs are rich in FR-Is. Therefore, we selected FR-Is from \citet{Massaro}.
If RGs listed in \citet{Mingo} or \citet{Massaro} is also listed in 4FGL-DR2, we define them as GeV-loud RGs.
Consequently, our sample contains 36 GeV-loud RGs (19 FR-I, 12 FR-II and 5 CSS) and 32 GeV-quiet RGs (11 FR-I, 20 FR-II, and 1 CSS); a total of 68 RGs. Our sample is given in Table \ref{sample}.
Figure \ref{hist} shows distributions of redshift and radio flux at 178 MHz for our sample RGs. Both of GeV-loud and GeV-quiet RGs have a similar distribution.

Black hole masses in Table \ref{sample} are quoted from literatures if available or estimated from $K_s$-Band and $B$-Band magnitudes if not previously reported. The formula used to estimate a BH mass $M_{\rm BH}$ from the $K_s$-Band magnitude $K_s$ is $\log\left({M\rm{_{BH}}/M_\odot}\right)=-0.45K_s-2.5$ (\citet{Dong}), and that used to estimate from $B$-Band magnitude $M_{\rm B}$ is $\log\left({M\rm{_{BH}}/M_\odot}\right)=-1.58-0.488M_{\rm B}$ \citep{Marchesini}.

\subsection{X-ray Data}
X-ray data used in this study are obtained from the {\it XMM-Newton}, {\it Chandra}, {\it Swift} and {\it NuSTAR} satellites. The priority of the satellite data selection was {\it XMM-Newton}, {\it Chandra}, and {\it Swift} satellites in that order. The {\it NuSTAR} satellite data were used if the X-ray spectra suffer from significant absorption. For multiple observations of the same object by the same satellite, the data with the highest photon count were used for {\it XMM-Newton}; moreover, the data with the longest observation time were used for the other three satellites.

The {\it XMM-Newton} data were reprocessed and analyzed using Science Analysis System (SAS) version 19.0.0 and the latest calibration data CCF (Current Calibration Files) as of April 18, 2021. {\it Chandra} data were calibrated using version 4.12 of CIAO (Chandra Interactive Analysis of Observations) and version 4.9.3 of CALDB (calibration database), and spectra were created using {\tt specextract}. {\it Swift} data were calibrated using {\tt XRTPIPELINE} version 0.13.5 and CALDB version 20200724 of High Energy Astrophysics Science Archive Research Center (HEASARC). Spectral files and ancillary response files were created using {\tt xselect}. For {\it Nustar} data, data were calibrated using {\tt NUPIPELINE} version 0.4.8 and version 20200912 of CALDB, and spectra were created using {\tt nuproducts}.

\section{Spectral Fitting and Result}
\subsection{Spectral Fitting}

Spectral fitting was performed using Xspec ver.12.11.1. If pile-up occurred in the {\it Chandra} data, we used Sherpa ver.2 to correct the pile-up and performed the analysis. If the pile-up rate was larger than $10\%$, the data were not used, and we used data from {\it Nustar} and {\it Swift}.

We set the minimum count for one bin to be one and performed fitting using the {\it C}-statistic. When correcting the Chandra satellite data for pile-up with Sherpa, we used a minimum count of 20 in one bin and performed the fitting using the $\chi^2$ test. We confirmed for all objects that the spectral fitting was good with a reduced $\chi^2$ value of $<1.3$ and no residual features are seen.

The components of the model used in this study are {\tt phabs}, {\tt zphabs}, {\tt pegpwrlw}, {\tt zgauss}, and {\tt apec}. {\tt phabs} represents the photoelectric absorption by our Galaxy. Hydrogen column density ($N\rm{_H}$) of this component was referred to the HESOFT command {\tt nh} which uses the HI4PI collaboration 2016 database. In the following analyses, when a model is redshift-dependent, the redshift is fixed to the value in Table \ref{sample}. {\tt zphabs} represents the absorption in the parent galaxy.  {\tt pegpwrlw} represents the power-law model typically found in AGN. The normalization of this component is the flux in the specified energy range. {\tt zgauss} is the Gaussian function to represent the iron emission line.  {\tt apec} is a hot plasma emission model. This model is used when there is high-temperature plasma around the AGN. The metal abundances of this component were fixed to 0.5 solar. {\tt const} component was introduced to fit spectra of the {\it XMM-Newton} and {\it Nustar} satellites, which have multiple detectors, to adjust relative normalizations. {\tt zedge} was introduced for Fornax A to show the absorption edge.

By combining of these model components, spectra were fitted by the following model A$\sim$H. Model A is a single power-law model as {\tt phabs*pegpwrlw}. Model B is a power-law model with absorption by the parent galaxy as {\tt phabs*zphabs*pegpwrlw}. Model C as {\tt phabs*(zphabs*pegpwrlw+zgauss)} has a gaussian in addition to the model B. Model D as {\tt phabs*(pegpwrlw+zphabs*pegpwrlw)} is a double power-law model; one of two power-law component is absorbed by the torus. Model E as {\tt phabs*(pegpwrlw+zphabs* pegpwrlw+zgauss)} has a gaussian in addition to the model D. Model F as {\tt phabs*(apec+pegpwrlw)}, model G as {\tt phabs*(apec+zphabs*pegpwrlw)}, and model H as {\tt phabs*(apec+pegpwrlw+zphabs*pegpwrlw)} have a high-temperature plasma component in addition to the model A, B, and D. Here, the models with two {\tt pegpwrlw} components has a common photon index. Figure \ref{mode1a-h} shows an example of spectral fitting by each model.
For objects which do not require an intrinsic absorption, an upper limit of intrinsic $N\rm{_H}$ was estimated by multiplying the unabsorbed power-law (best-fit model) by an additional {\tt zpha}.

\subsection{Result}
Table \ref{result} summarizes the best fit parameters obtained based on the spectral analysis. Error for each parameter is the 90\% confidence interval for one parameter. Table \ref{result} shows that there are four sources with the iron emission line and 10 sources with the {\tt apec} component. Table \ref{Edington} shows the X-ray luminosity $L_{\rm X}$ and Eddington luminosity ratio for the power-law component. The luminosity of the power-law component is corrected for absorption. The 2-10 keV X-ray luminosity is in the range of  $1.35\times10^{40}$--$9.19\times10^{45}\,\rm{erg\,s^{-1}}$. For the 64 objects whose BH mass was available, the estimated Eddington luminosity ratio for the X-ray power-law emission is mostly in the range of $6.54\times10^{-7}$--1.76.

Figures \ref{nonefermispec} and \ref{fermispec} show an example of spectra of GeV-quiet and GeV-loud RGs, respectively. We can see that most of GeV-loud RGs do not show a significant absorption in the soft X-ray band; however, many GeV-quiet RGs suffer from  heavy absorption. Table \ref{absratio} shows a fraction of sources with absorption $N\rm{_H}$ greater than $10^{22}\rm{cm}^{-2}$ for GeV-loud and GeV-quiet RGs. From this table, we can see that few GeV-loud RGs undergo absorption, while $\sim50\%$ of GeV-quiet RGs undergo absorption. Table 4 shows a fraction for each of RG classifications. The fraction of sources with heavy absorption for FR-II RGs is as high as 40\%, while that of others are $\leq$20\%.

\section{Discussion}
\subsection{Absorption $N\rm{_H}$ and Gamma-ray loudness}
We reported that $\sim50\%$ of the GeV-quiet RGs have a large absorption column density $N\rm{_H}$, while few GeV-loud RGs have a large $N\rm{_H}$. The radiation from the center of AGNs is absorbed by the torus depending on the viewing angle. Therefore, the absorption is large when the torus is viewed from the side, while the absorption is small when it is viewed from a small inclination angle to the torus axis. Because the jet and torus axis are the same, we can assume that the viewing angle of the jet is smaller as the absorption becomes smaller. The jets of RGs could be brighter in gamma rays when the jet viewing angle is small and thus the beaming effect is large. This suggests that the GeV-loud RGs are observed from a relatively small viewing angle to the jet axis. When the jet is seen from a small angle, it is suggested that the X-ray radiation from the center of AGNs is not blocked by the torus and thus the absorption is small.

On the other hand, GeV-quiet RGs can be considered to have a weak jet beaming effect and thus gamma-ray radiation is faint, making them difficult to detect with Fermi. Therefore, we can assume that the GeV-quiet RGs are observed from a large viewing angle to the jet axis. In that case, we observe the center of AGNs through the torus. Therefore, it can be understood that the X-ray emission from the center of AGNs is easily blocked by the torus and most of the GeV-quiet RGs show absorbed spectra.

\subsection{Absorption $N\rm{_H}$ and X-ray luminosity $L\rm{_x}$}
\begin{figure}[H]
  \centering
  \includegraphics[width=12cm]{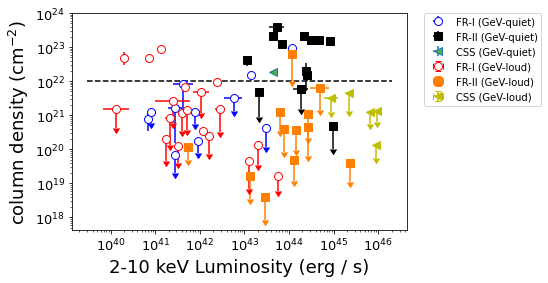}
  \caption{X-ray luminosity $L\rm{_x}$ vs Absorption column density $N\rm{_H}$. Blue circles, black squares, green triangles, red circles, orange squares and yellow triangles represent GeV-quiet FR-I, GeV-quiet FR-II, GeV-quiet CSS, GeV-loud FR-I, GeV-loud FR-II and GeV-loud CSS, respectively.}
  \label{LNH}
\end{figure}

From Tables \ref{result} and \ref{Edington}, we created a scatter plot of the absorption column density $N\rm{_H}$ and X-ray luminosity (Figure \ref{LNH}). This figure shows that $\sim50\%$ of the medium-luminosity ($10^{43}-10^{45}$ erg s$^{-1}$ RGs show a large $N\rm{_H}$, while few of the low and high luminosity RGs show a large $N\rm{_H}$. The previous studies of the relationship between absorption and luminosity of RGs (Panessa\,et\,al.\,2016, Kuraszkiewicz\,et\,al.\,2021) show the relation of mid- and high-luminosity RGs, and our results are consistent with these studies. Moreover, our result on low-luminosity RGs is novel.

In Seyfert galaxies, as shown by \citet{Beckmann} and \citet{Burlon}, $\sim50\%$ of the medium-luminosity sources show a large $N\rm{_H}$, and few low and high luminosity sources show a large $N\rm{_H}$. This suggests that the absorption-luminosity relation for RGs, including low-luminosity RGs, follows the same trend as that for Seyfert galaxies. Because the fraction of AGNs with a large $N\rm{_H}$ at high luminosities is small, the dust torus is not formed with a large solid angle or a large thickness for brighter AGNs. This can be attributed to a physical process in which the high radiation flux from the center of AGNs affects the physical structure of the torus \citep{Ueda}. Similarly, the fraction of AGNs with a large $N\rm{_H}$ at low luminosities is small. It is suggseted that the torus is not formed in low-luminosity ANGs (e.g., \cite{Burlon}). The same scenario can be applied to RGs, and thus they may not have a torus at low luminosities. Furthermore, as reported by \citet{Fukazawa}, the X-ray emission of low-luminosity RGs could be synchrotron radiation from the jet. In that case, the emission is not absorbed by the torus.

\subsection{Photon index $\Gamma\rm{_x}$ and Eddington ratio $L\rm{_x}/{\it L}\rm{_{Edd}}$}
\begin{figure}[H]
  \centering
  \includegraphics[width=12cm]{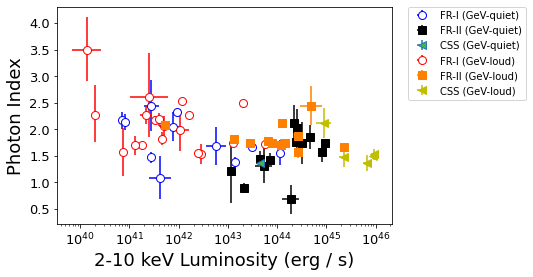}
  \caption{X-ray luminosity $L\rm{_x}$ vs photon index $\Gamma\rm{_x}$. Markers are the same as figure \ref{LNH}}
  \label{GL}
\end{figure}

\begin{figure}[H]
  \centering
  \includegraphics[width=12cm]{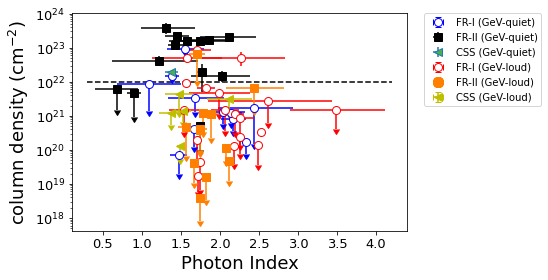}
  \caption{Photon index $\Gamma\rm{_x}$ vs Absorption column density $N\rm{_H}$. Markers are the same as figure \ref{LNH}}
  \label{GNH}
\end{figure}

Figures \ref{GL} and \ref{GNH} are a scatter plot of X-ray luminosity as well as power-law index and that of the power-law index and absorption column density $N\rm{_H}$. These figures demonstrated no dependence of photon index on $N\rm{_H}$ and X-ray luminosity. The average of photon index is $1.83\,(0.46)$, $1.94\,(0.42)$ and $1.70\,(0.46)$ for all RGs, GeV-loud RGs and GeV-quiet RGs, respectively, where values in the parentheses are standard deviation. It is $1.88\,(0.47)$ and $1.70\,(0.42)$ for RGs with NH$<10^{22}\rm{cm}^{-2}$ and those with NH$>10^{22}\rm{cm}^{-2}$, respectively.

\citet{Sambruna}, reported weak evidence that Broad Line Radio Galaxies (BLRGs) have a flatter X-ray spectrum than that of the radio-quiet Seyfert 1 galaxies. \citet{Kang} reported that the power-law photon index of RGs is flatter than that of radio-quiet AGNs. They analyzed the {\it NuSTAR} data of non-Compton-thick RGs and compared them with the results of \cite{Panagiotou} on radio-quiet AGNs. The average of photon index in \citet{Kang} is $1.73\,(0.15)$ and $1.90\,(0.21)$ for RGs and radio-quiet AGNs, respectively.

Photon index $1.83\,(0.46)$ for all RGs in our result is steeper than the photon index for RGs in \citet{Kang}. Because the photon index of the GeV-loud RGs appears to be larger, we performed a KS test on the distribution of the photon index of GeV-loud RGs and that of GeV-quiet RGs and obtained a p-value of 0.3, indicating that the GeV-loud and GeV-quiet RGs have a significantly different photon index distribution. If the high-energy tail of synchrotron radiation from the jet is observed in the X-ray band, the photon index is expected to be larger \citep{Fukazawa}; therefore, the photon index of GeV-loud RGs could be affected by the jet. The photon index of the GeV-quiet RGs may be flatter than that of the radio-quiet AGN \citep{Kang}; however, because $\sim$half of the GeV-quiet RGs suffer from the absorption, this flatness might be caused by the systematic effect from absorption.

Photon index could depend on the Eddington luminosity ratio. Therefore, we investigated the relation between photon index and Eddington luminosity ratio; however, we do not identify any dependence within errors.

\begin{ack}
The authors thanks to the editor and the anonymous referee for their helpful comments to improve the paper.
This research made use of the XMM-Newton, Chandra, Swift and NuSTAR Data, and also the NASA/IPAC Extragalactic Database (NED).
\end{ack}




\begin{table}
\caption{Sample of Radio Galaxies}
\label{sample}
\centering
\begin{tabular}[H]{clcccccccc}
\hline
No.&Source$\rm{^{a}}$&Data$\rm{^{b}}$&ObsID&Redshift&$N\rm{_H}\rm{^{c}}$&F$_{178}\rm{^{d}}$&class$\rm{^{e}}$&$M\rm{_{BH}}\rm{^{f}}$&Reference$\rm{^{g}}$\\
&&&&&$10^{20}\,\rm{cm^{-2}}$&Jy&&$10^8M_\odot$&\\
\hline
\multicolumn{10}{c}{GeV-quiet RGs}\\
\hline
1&3C\,433&{\it Chandra}&7881&0.10160&8.21&57.5&FR-I&16.5&2,\,1\\
2&PKS\,0409-75&{\it XMM-Newton}&0651281901&0.69300&7.92&55.9&FR-II&10.5$\rm{^{i}}$&1,\,7\\
3&PKS\,2356-61&{\it XMM-Newton}&0677180601&0.09631&1.33&51.5&FR-II&9.10&1,\,1\\
4&PKS\,1814-63&{\it XMM-Newton}&0146340601&0.06466&6.01&43.7&CSS&6.80&1,\,1\\
5&PKS\,0442-28&{\it Swift}&80964002&0.14700&2.46&41.8&FR-II&13.3&1,\,1\\
6&PKS\,1559+02&{\it Chandra}&6841&0.10480&5.70&36.3&FR-II&13.8&1,\,1\\
7&3C\,459&{\it XMM-Newton}&0651280101&0.22012&5.30&29.2&FR-II&19.5$\rm{^{h}}$&1,\,4\\
8&3C\,227&{\it Nustar}&60061329002&0.08627&1.99&28.6&FR-II&8.40&1,\,1\\
9&PKS\,1938-15&{\it XMM-Newton}&0671871001&0.45200&7.40&25.8&FR-II&-&1,\,1\\
10&3C\,403&{\it Nustar}&60061293002&0.05900&13.2&25.2&FR-II&9.20&1,\,1\\
11&3C\,032&{\it XMM-Newton}&0651281001&0.40000&1.60&25.1&FR-II&19.6&1,\,1\\
12&3C\,445&{\it Nustar}&60160788002&0.05588&4.80&24.6&FR-II&11.8&1,\,1\\
13&3C\,327.1&{\it XMM-Newton}&0651281201&0.46200&6.89&23.4&FR-II&-&1,\,-\\
14&PKS\,0039-44&{\it XMM-Newton}&0651280901&0.34600&2.63&23.2&FR-II&14.6&1,\,1\\
15&3C\,062&{\it Chandra}&10320&0.14700&1.62&21.9&FR-II&10.3&1,\,1\\
16&PKS\,0349-27&{\it Chandra}&11497&0.06569&0.737&21.5&FR-II&3.60&1,\,1\\
17&PKS\,0043-42&{\it Chandra}&10319&0.05280&1.12&21.0&FR-II&9.40&1,\,1\\
18&PKS\,1954-55&{\it Chandra}&11505&0.05810&4.46&16.5&FR-I&33.5&1,\,4\\
19&3C\,015&{\it Chandra}&17128&0.07338&2.19&16.4&FR-I&5.96$\rm{^{i}}$&2,\,5\\
20&3C\,031&{\it XMM-Newton}&0551720101&0.01700&5.22&16.3&FR-I&8.20&2,\,1\\
21&3C\,105&{\it Nustar}&60261003002&0.08900&10.4&16.2&FR-II&4.10&1,\,1\\
22&3C\,029&{\it Chandra}&12721&0.04503&3.19&15.7&FR-I&7.63$\rm{^{i}}$&2,\,5\\
23&PKS\,1547-79&{\it XMM-Newton}&0651281401&0.48300&10.3&12.4&FR-II&34.4&1,\,1\\
24&3C\,346&{\it Chandra}&3129&0.16201&4.99&12.0&FR-I&17.0&2,\,1\\
25&PKS\,0347+05&{\it XMM-Newton}&0651280801&0.33900&12.8&11.8&FR-II&32.8&1,\,1\\
26&3C\,130&{\it Chandra}&14957&0.10900&36.0&10.8&FR-I&20.8$\rm{^{h}}$&2,\,4\\
27&3C\,296&{\it XMM-Newton}&0201860101&0.02470&1.84&10.4&FR-I&12.3&2,\,1\\
28&PKS\,1306-09&{\it XMM-Newton}&0671871201&0.46685&3.01&9.79&FR-II&33.3&1,\,1\\
29&3C\,402&{\it XMM-Newton}&0823050101&0.02595&11.9&9.55&FR-I&1.81$\rm{^{i}}$&2,\,5\\

\end{tabular}
\end{table}
\begin{table}
\centering
\begin{tabular}[t]{clcccccccc}
\multicolumn{10}{l}{\textbf{Table \ref{sample}.}\,(Continued)}\\
\hline
    No.&Source$\rm{^{a}}$&Data$\rm{^{b}}$&ObsID&Redshift&$N\rm{_H}\rm{^{c}}$&F$_{178}\rm{^{d}}$&class$\rm{^{e}}$&$M\rm{_{BH}}\rm{^{f}}$&Reference$\rm{^{g}}$\\
    &&&&&$10^{20}\,\rm{cm^{-2}}$&Jy&&$10^8M_\odot$&\\
\hline
30&PKS\,0620-52&{\it XMM-Newton}&0852980601&0.05110&3.71&8.96&FR-I&22.7&1,\,1\\
31&PKS\,1934-63&{\it XMM-Newton}&0784610201&0.18129&5.83&7.37&FR-II&10.9&1,\,1\\
32&3C\,213.1&{\it Chandra}&9307&0.19392&2.57&6.60&FR-I&4.04$\rm{^{h}}$&2,\,4\\
    \hline
   \multicolumn{10}{c}{GeV-loud RGs}\\
\hline
 
33&Pictor\,A&{\it XMM-Newton}&0206390101&0.03501&3.62&441&FR-II&1.21$\rm{^{h}}$&3,\,4\\
34&Fornax\,A&{\it XMM-Newton}&0502070201&0.00591&1.99&273&FR-I&0.544$\rm{^{h}}$&3,\,4\\
35&PKS\,2153-69&{\it XMM-Newton}&0152670101&0.02827&2.49&81.7&FR-II&1.39$\rm{^{h}}$&3,\,4\\
36&PKS\,0521-36&{\it XMM-Newton}&065760201&0.05655&3.27&71.7&FR-II&11.4$\rm{^{h}}$&3,\,4\\
37&3C\,380&{\it Swift}&81221001&0.69000&6.29&65.5&CSS&27.8$\rm{^{i}}$&3,\,5\\
38&Cen\,B&{\it XMM-Newton}&092140101&0.01292&68.0&50.0&FR-I&0.988$\rm{^{h}}$&3,\,4\\
39&NGC\,4261&{\it XMM-Newton}&0502120101&0.00726&1.61&44.1&FR-I&0.460$\rm{^{h}}$&3,\,4\\
40&3C\,264&{\it XMM-Newton}&0602390101&0.02160&1.87&25.3&FR-I&5.60&3,\,1\\
41&3C\,286&{\it Chandra}&15006&0.84989&1.11&25.2&CSS&3.16&3,\,6\\
42&3C\,309.1&{\it Chandra}&3105&0.90500&2.89&24.2&CSS&42.6$\rm{^{i}}$&3,\,5\\
43&PKS\,1839-48&{\it Chandra}&10321&0.11080&5.25&23.4&FR-I&21.1&1,\,1\\
44&PKS\,1821-327&{\it XMM-Newton}&0650591401&0.35500&12.4&23.3&FR-II&-&3,\,-\\
45&PKS\,0625-35&{\it XMM-Newton}&0302440601&0.05459&6.36&21.6&FR-I&12.9&1,\,1\\
46&3C\,216&{\it Chandra}&15002&0.67002&1.34&21.2&CSS&0.100&3,\,6\\
47&3C\,138&{\it Chandra}&14996&0.76000&20.1&21.0&CSS&6.30&3,\,1\\
48&PKS\,2324-02&{\it XMM-Newton}&0405860101&0.18873&4.49&11.5&FR-II&13.5$\rm{^{h}}$&3,\,4\\
49&3C\,303&{\it Nustar}&60463048002&0.14121&1.68&11.3&FR-II&3.18$\rm{^{i}}$&3,\,5\\
50&NGC\,2484&{\it XMM-Newton}&0602390101&0.04083&4.74&9.35&FR-II&2.39$\rm{^{h}}$&3,\,4\\
51&B2\,1113+29&{\it XMM-Newton}&0550270101&0.04868&1.19&7.44&FR-I&1.92$\rm{^{i}}$&3,\,5\\
52&PKS\,2300-18&{\it Swift}&31729002&0.12893&1.90&5.48&FR-II&10.4$\rm{^{h}}$&3,\,4\\
53&NGC\,315&{\it XMM-Newton}&0305290201&0.01648&5.88&5.20&FR-I&0.951$\rm{^{h}}$&3,\,4\\
54&4C\,+40.01&{\it Chandra}&5669&0.25500&5.93&5.10&FR-II&12.5$\rm{^{h}}$&3,\,4\\
55&PKS\,1514+00&{\it XMM-Newton}&0103860601&0.05259&4.14&3.90&FR-I&-&3,\,-\\
56&NGC\,6328&{\it XMM-Newton}&0804520301&0.01443&5.90&2.97&FR-I&0.968$\rm{^{h}}$&3,\,4\\
57&4C\,+39.12&{\it Chandra}&857&0.02020&12.5&2.70&FR-I&1.61$\rm{^{h}}$&3,\,4\\
58&TXS\,1516+064&{\it XMM-Newton}&018741001&0.10208&2.91&1.14&FR-II&6.17$\rm{^{h}}$&3,\,4\\
\end{tabular}
\end{table}
\begin{table}
\centering
\begin{tabular}[th]{clcccccccc}
\multicolumn{10}{l}{\textbf{Table \ref{sample}.}\,(Continued)}\\
\hline
    No.&Source$\rm{^{a}}$&Data$\rm{^{b}}$&ObsID&Redshift&$N\rm{_H}\rm{^{c}}$&F$_{178}\rm{^{d}}$&class$\rm{^{e}}$&$M\rm{_{BH}}\rm{^{f}}$&Reference$\rm{^{g}}$\\
    &&&&&$10^{20}\,\rm{cm^{-2}}$&Jy&&$10^8M_\odot$&\\
    \hline
59&TXS\,0149+710&{\it Swift}&10149002&0.02281&30.7&1.06&FR-I&1.06$\rm{^{h}}$&3,\,4\\
60&IC\,1531&{\it XMM-Newton}&0202190301&0.02564&2.41&1.05&FR-I&1.44$\rm{^{h}}$&3,\,4\\
61&NGC\,2329&{\it Chandra}&5900&0.01933&7.49&0.943&FR-I&1.46$\rm{^{h}}$&3,\,4\\
62&IC\,310&{\it XMM-Newton}&0151560101&0.01885&11.4&0.732&FR-I&1.29$\rm{^{h}}$&3,\,4\\
63&B2\,1447+27&{\it Swift}&40619002&0.03059&2.89&0.600&FR-I&2.66$\rm{^{i}}$&3,\,5\\
64&PKS\,2331-240&{\it XMM-Newton}&0760990101&0.04770&1.65&0.577&FR-I&1.18$\rm{^{h}}$&3,\,4\\
65&NGC\,2892&{\it Chandra}&18038&0.02246&4.19&0.566&FR-I&1.41$\rm{^{h}}$&3,\,4\\
66&B3\,0309+411B&{\it XMM-Newton}&0306680301&0.13400&10.1&0.547&FR-II&7.49$\rm{^{h}}$&3,\,4\\
67&NGC\,3894&{\it Chandra}&10389&0.01077&1.83&0.371&FR-I&0.589$\rm{^{h}}$&3,\,4\\
68&B3\,1009+427&{\it Swift}&85352004&0.36491&1.58&0.177&FR-II&27.8$\rm{^{h}}$&3,\,4\\
   \hline
   \multicolumn{10}{l}{a: Name of objects. b: Satellite of X-ray data. c: Galactic hydrogen column density (HI4PI collaboration 2016).}\\
   \multicolumn{10}{l}{d: Radio flux at 178MHz. e: Classification of RGs. f: BH masses. (h) BH mass estimated using $K_s$-band data.}\\
   \multicolumn{10}{l}{(i) BH mass estimated using $B$-band data. g: Left and right are references for object selection and BH mass,}\\
   \multicolumn{10}{l}{respectively. 1: \citet{Mingo} 2: \citet{Massaro} 3: 4FGL-DR2 catalog}\\
   \multicolumn{10}{l}{4: $K_s$-Band~magnitude (2MASS catalog) 5: \citet{Marchesini} 6: \citet{Yuan} 7: \citet{Massardi}}\\
\end{tabular}
\end{table}

\begin{figure}[ht]
  \begin{minipage}[h]{0.5\hsize}
    \centering
    {(a) PKS\,0625-35~(model\,A)}
    \includegraphics[height=7.5cm,width=9cm]{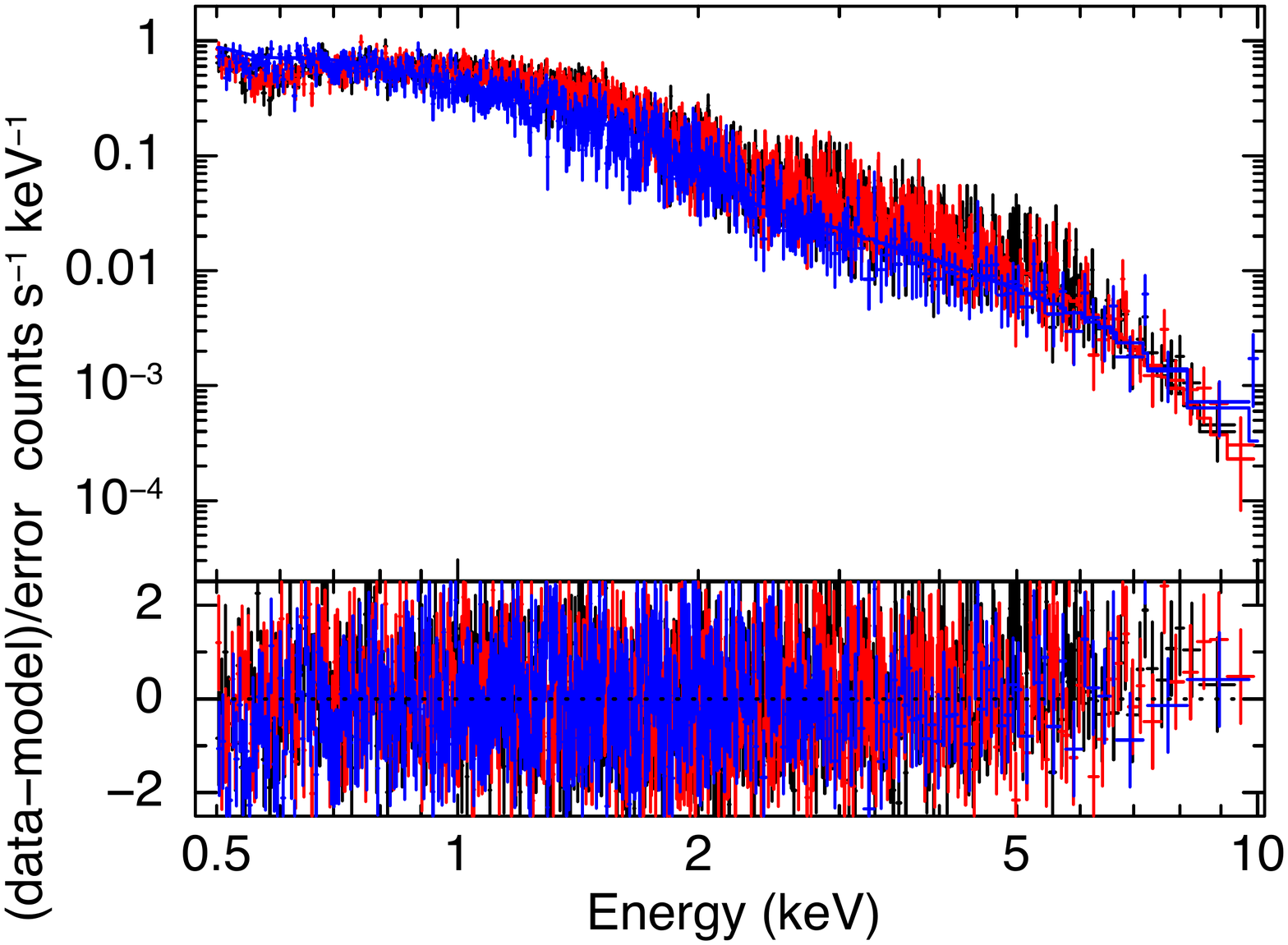}
    \label{modela}
    \end{minipage} 
  \begin{minipage}[h]{0.5\hsize}
    \centering
    {(b) 3C\,227~(model\,B)}
    \includegraphics[height=7.5cm,width=9cm]{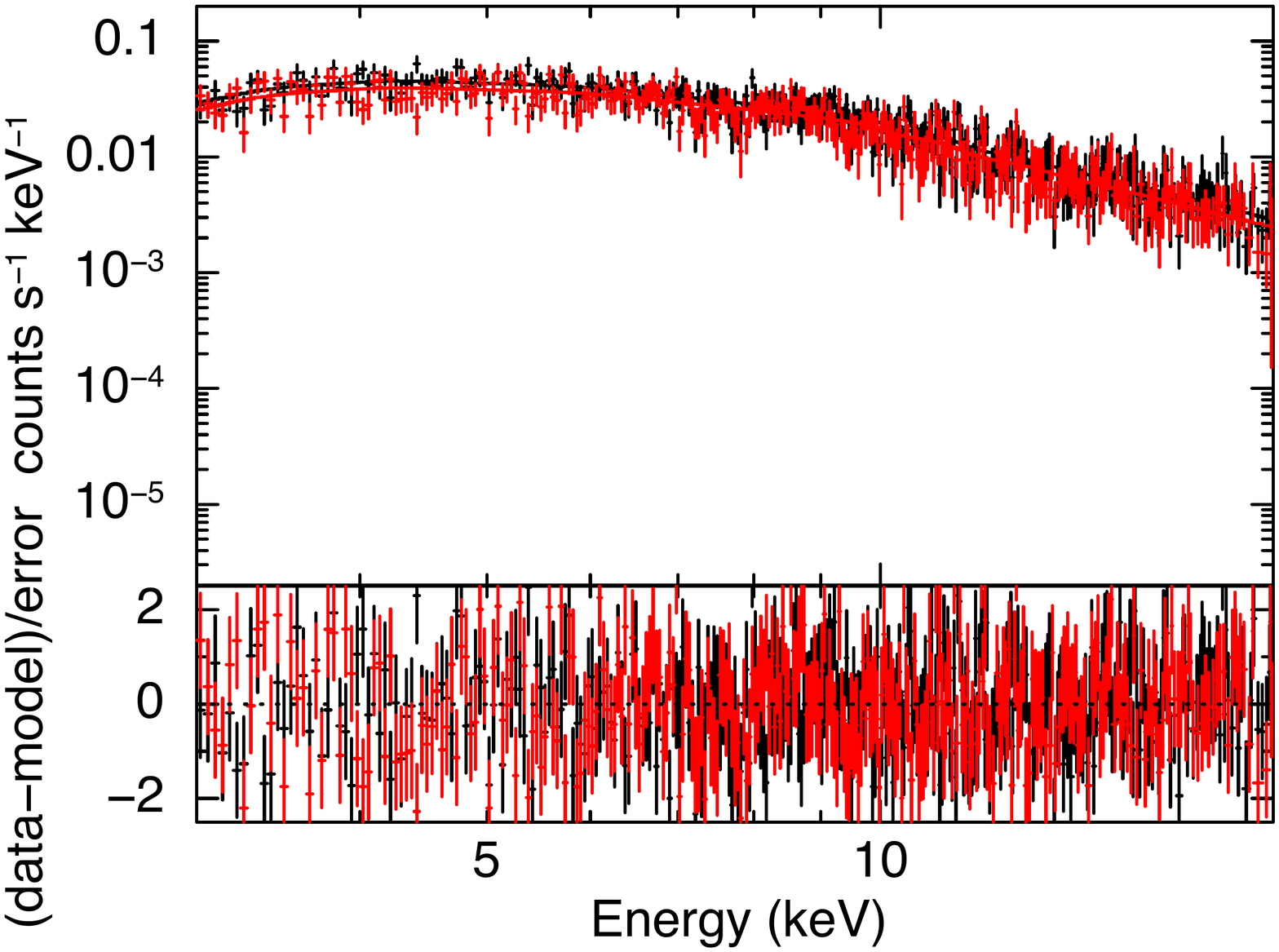}
    \label{modelb}
  \end{minipage} \\
        \begin{minipage}[h]{0.5\hsize}
        \centering
        {(c) 3C\,105~(model\,C)}
        \includegraphics[height=7.5cm,width=9cm]{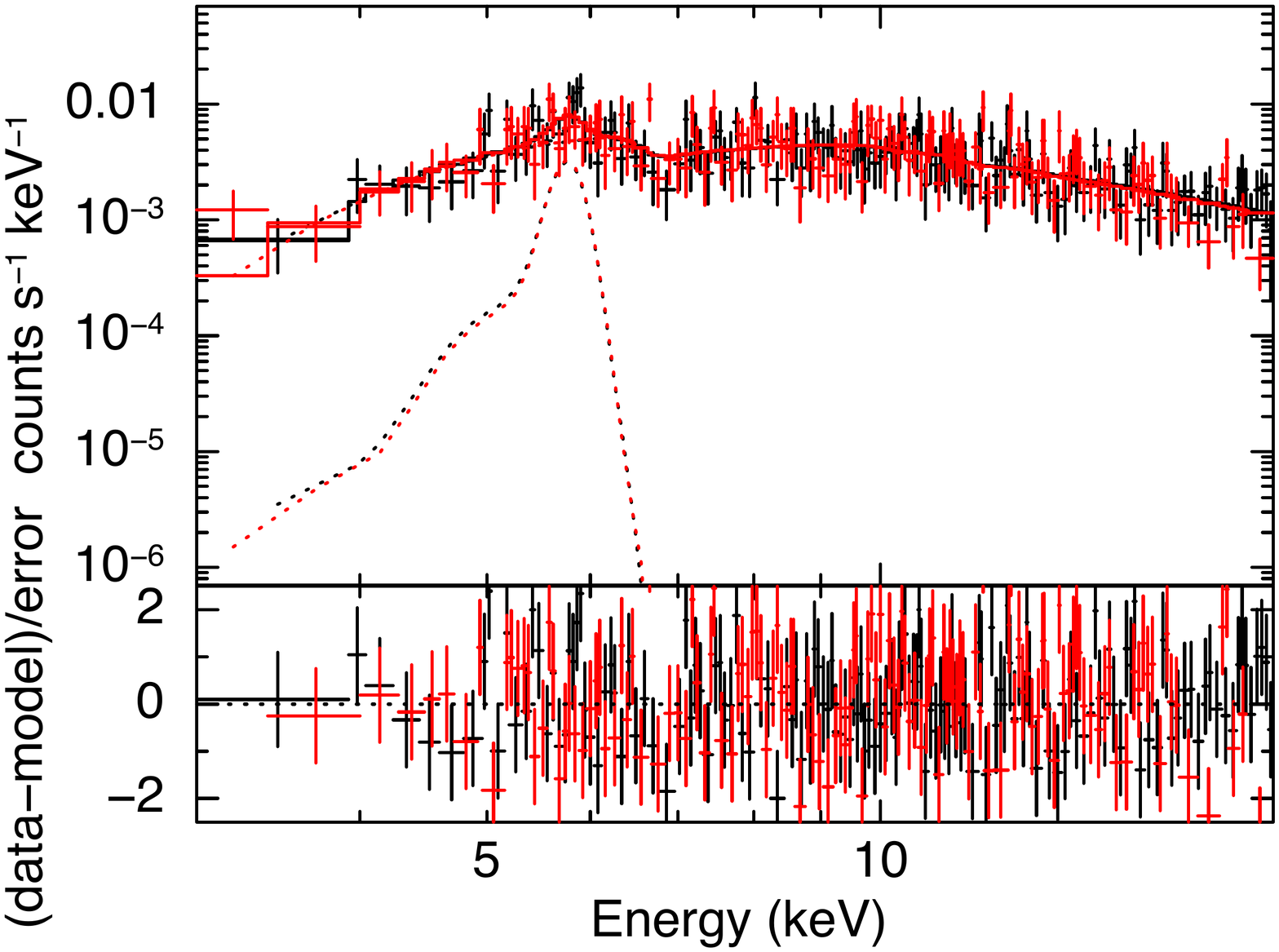}
        \label{modelc}
      \end{minipage}
      \begin{minipage}[h]{0.5\hsize}
        \centering
        {(d) 3C\,032~(model\,D)}
        \includegraphics[height=7.5cm,width=9cm]{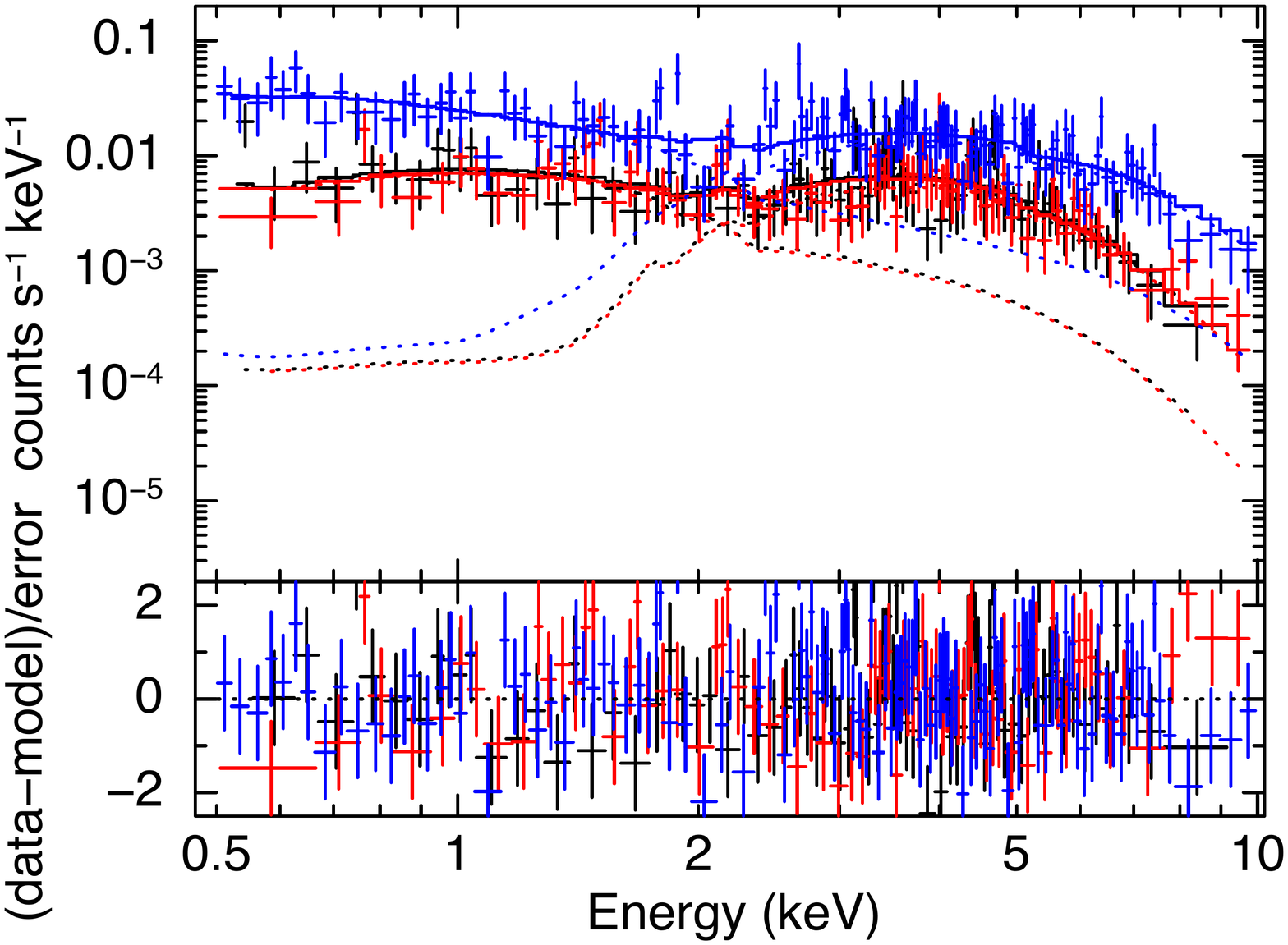}
        \label{modeld}
      \end{minipage}\\
      \caption{Example of spectra fitted with the Model A-H. Panel (a), (d), (e), (f), (g) and (h) are {\it XMM-Newton} data, where black, red, and blue symbols are MOS1, MOS2 and PN spectra, respectively. Panel (b) and (c) are {\it NuSTAR} data, where black and red symbols are FPMA and FPMB spectra, respectively.  The solid line represents the best-fit total model, while dotted lines are the model components. The bottom panels show the residuals in units of $\sigma$. (a) PKS\,0625-35 with model A ({\tt phabs*pegpwrlw}). (b) 3C\,227 with model B ({\tt phabs*zphabs*pegpwrlw}). (c) 3C\,105 with model C ({\tt phabs*(zphabs*pegpwrlw+zgauss)}). (d) 3C\,032 with model D ({\tt phabs*(pegpwrlw+zphabs*pegpwrlw)}).}
      \clearpage
     \end{figure}
\begin{figure}[ht]
    \addtocounter{figure}{-1}
     \begin{minipage}[h]{0.5\hsize}
        \centering
        {(e) PKS\,2356-61~(model\,E)}
        \includegraphics[height=7.5cm,width=9cm]{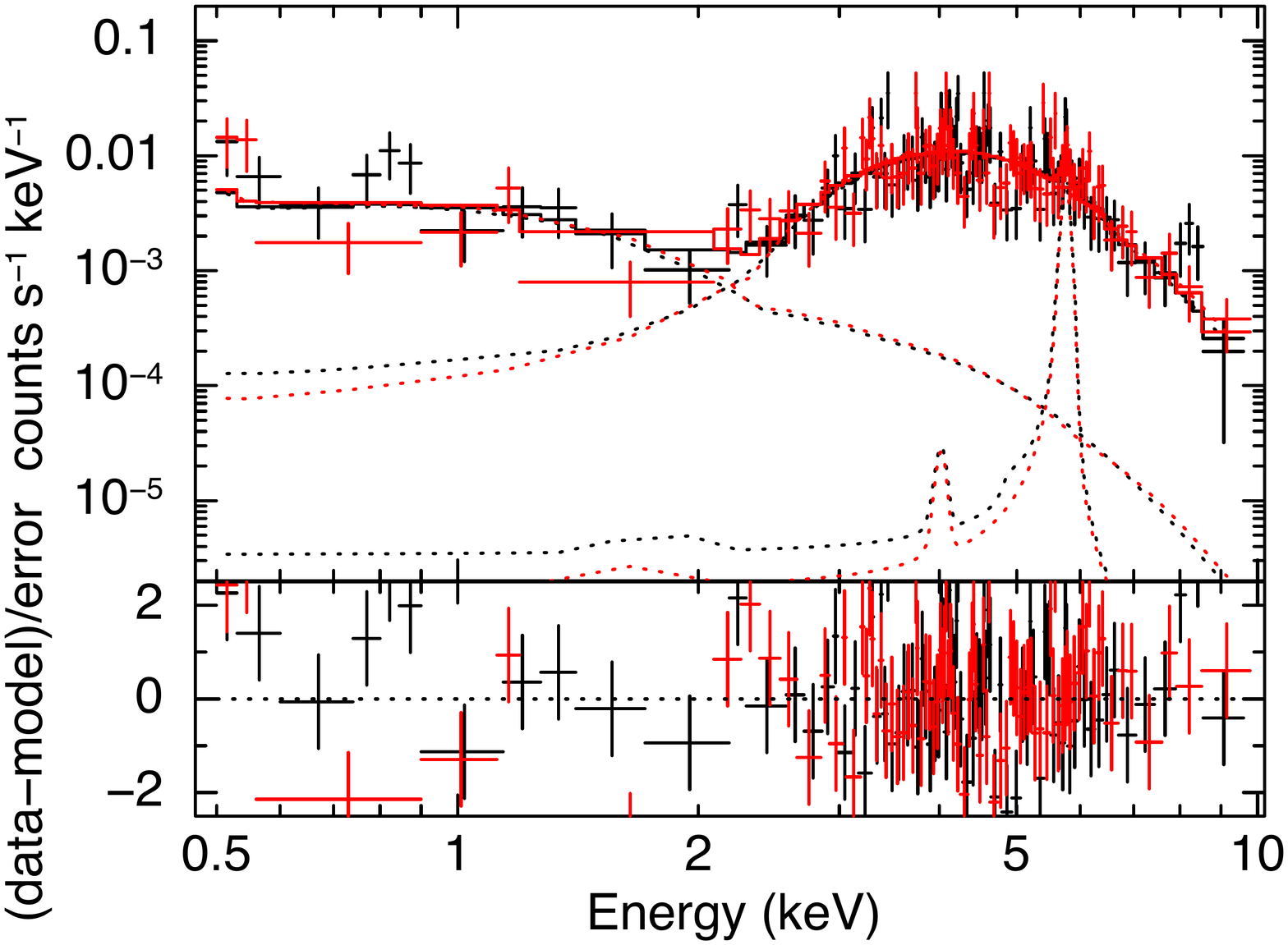}
        \label{modele}
      \end{minipage} 
      \begin{minipage}[h]{0.5\hsize}
        \centering
        {(f) NGC\,2484~(model\,F)}
        \includegraphics[height=7.5cm,width=9cm]{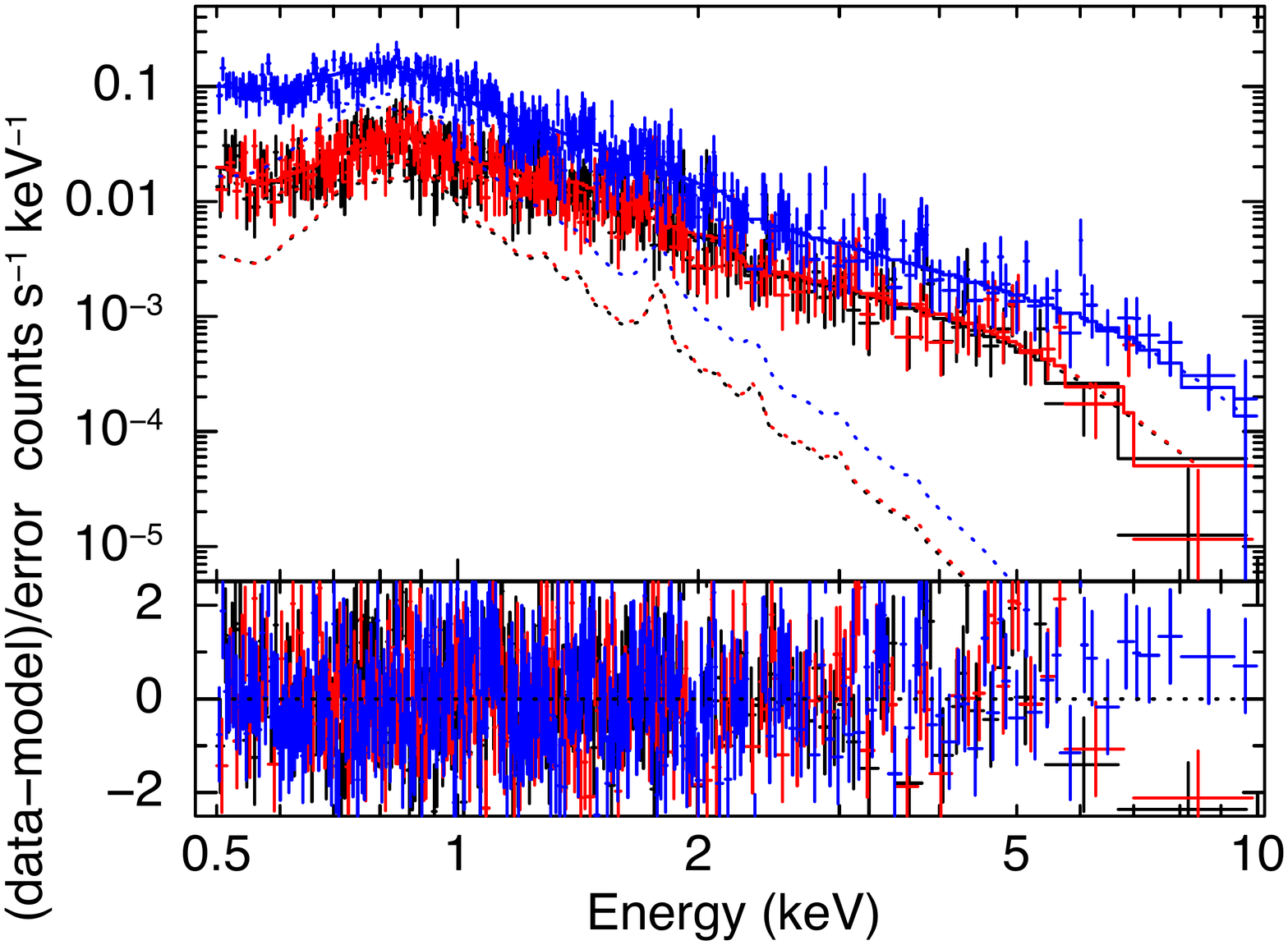}
        \label{modelf}
      \end{minipage} \\

      \begin{minipage}[h]{0.5\hsize}
        \centering
        {(g) 3C\,031~(model\,G)}
        \includegraphics[height=7.5cm,width=9cm]{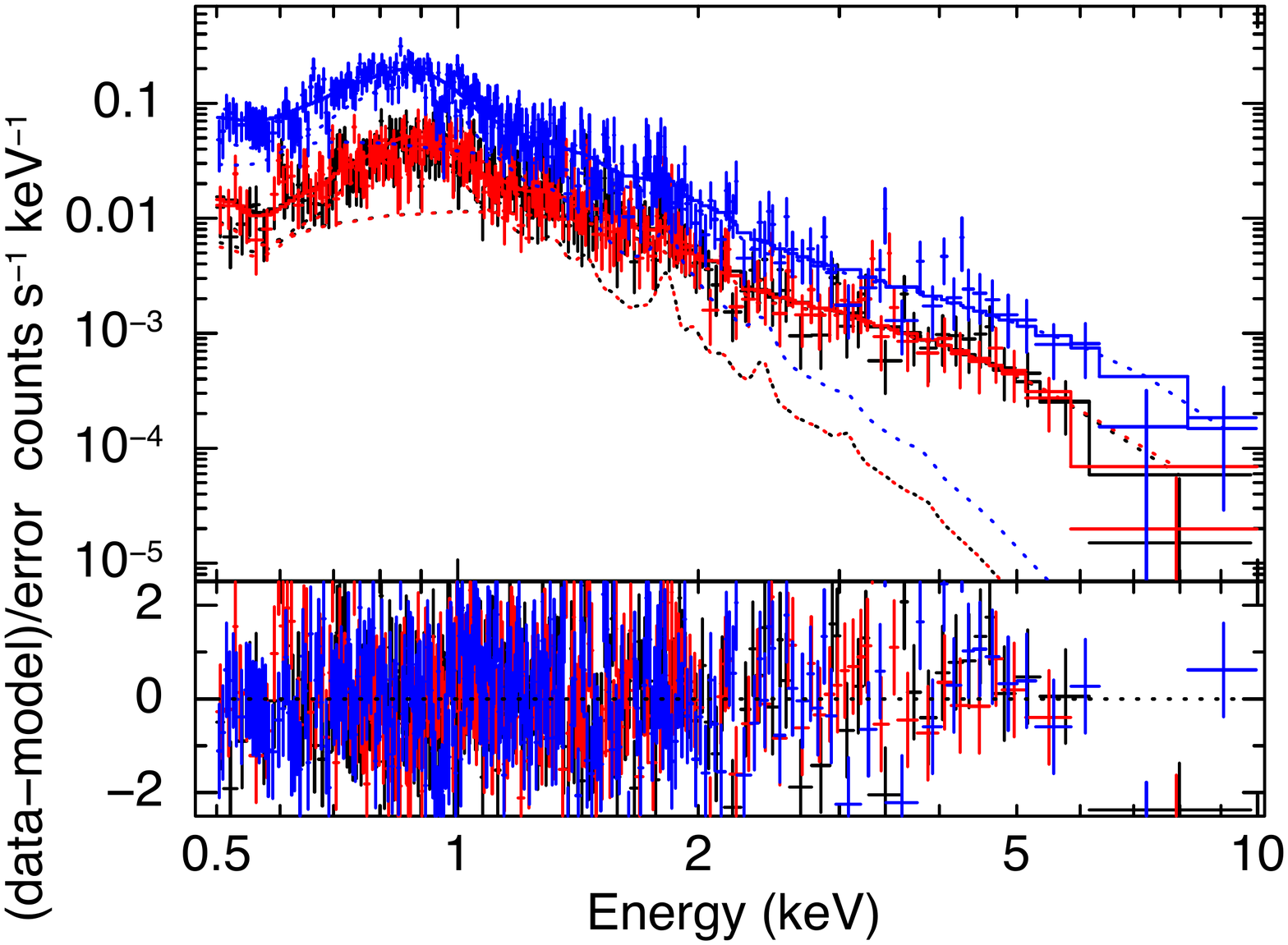}
        \label{modelg}
      \end{minipage} 
      \begin{minipage}[h]{0.5\hsize}
        \centering
        {(h) NGC\,4261~(model\,H)}
        \includegraphics[height=7.5cm,width=9cm]{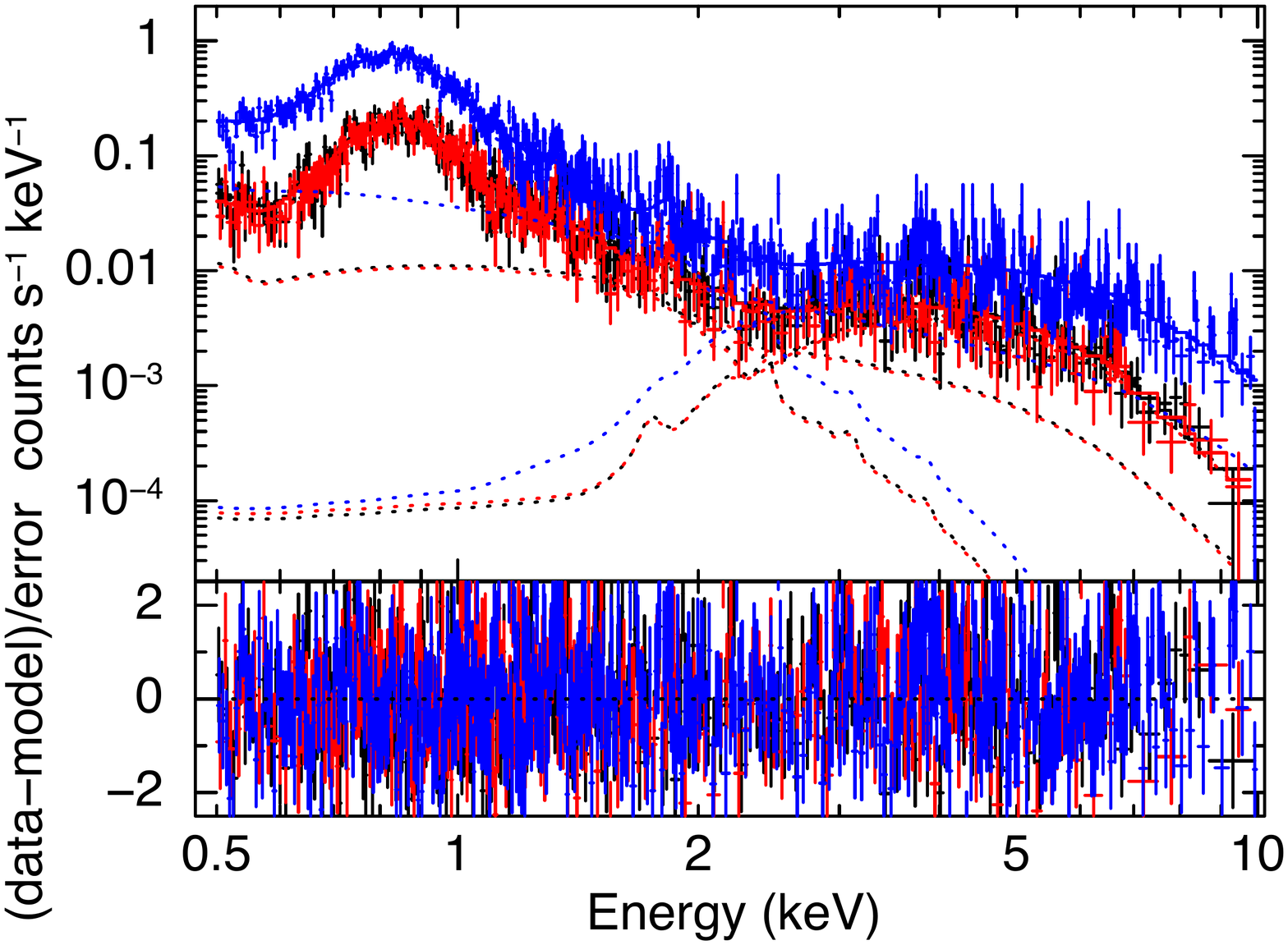}
        \label{modelh}
      \end{minipage} \\
     \caption{(Continued) (e) PKS\,2356-61 with model E ({\tt phabs*(pegpwrlw +zphabs* pegpwrlw+zgauss)}). (f) NGC\,2484 with model F ({\tt phabs*(apec+pegpwrlw)}). (G) 3C\,031 with model G ({\tt phabs*(apec+zphabs*pegpwrlw)}). (h) NGC\,4261 with model H ({\tt phabs*(apec+pegpwrlw+zphabs*pegpwrlw)}).}\label{mode1a-h}

\end{figure}

\begin{landscape}
 \begin{table}
    \caption{Summary of best-fit parameters of x-ray data spectral fitting}
    \label{result}
     \centering
\begin{tabular}[H]{ccccccccccc}
    \hline
   &&\multicolumn{2}{c}{apec}&\multicolumn{2}{r}{None absorbed pegpwrlw}&zphabs&\multicolumn{2}{r}{absorbed pegpwrlw}&\multicolumn{1}{c}{zgauss}&\\
   \hline
   No.&Model&{\it kT}&Norm&$\Gamma_u$&Norm1&Intrinsic~$N\rm{_H}$&$\Gamma_a$&Norm2&EW&${\it C}\rm{-stat}/\rm{d.o.f.}$\\
   &&keV&$10^{-5}$&&$10^{-12}\,\rm{erg\,cm^{-2}\,s^{-1}}$&$10^{22}\rm{cm^{-2}}$&&$10^{-12}\,\rm{erg\,cm^{-2}\,s^{-1}}$&eV&\\
   (1)&(2)&(3)&(4)&(5)&(6)&(7)&(8)&(9)&(10)&(11)\\

    \hline
1&D&-&-&$1.55^{+0.24}_{-0.23}$&$3.08^{+1.52}_{-1.02}\times10^{-2}$&$9.13^{+1.04}_{-0.96}$&$1.55^{+0.24}_{-0.23}$&$4.42^{+0.38}_{-0.29}$&-&136.5/116*\\
2&A&-&-&$1.99\pm0.07$&$2.46^{+0.31}_{-0.28}\times10^{-1}$&$<1.40\times10^{-1}$&-&-&-&1010.3/1156\\
3&E&-&-&$2.11^{+0.36}_{-0.35}$&$5.46^{+3.83}_{-2.30}\times10^{-2}$&$20.7^{+2.9}_{-2.8}$&$2.11^{+0.36}_{-0.35}$&$9.27^{+1.88}_{-1.46}$&$432^{+301}_{-237}$&672.0/755\\
4&C&-&-&-&-&$1.89^{+0.16}_{-0.15}$&$1.37^{+0.09}_{-0.08}$&$4.57^{+0.29}_{-0.25}$&$108^{+68}_{-54}$&2353.2/2830\\
5&D&-&-&$2.02^{+0.39}_{-0.37}$&$1.23^{+1.90}_{-0.86}\times10^{-1}$&$1.50^{+0.57}_{-0.50}$&$2.02^{+0.39}_{-0.37}$&$4.15^{+0.66}_{-0.58}$&-&192.1/260\\
6&E&-&-&$2.93\pm0.19$&$2.16^{+0.66}_{-0.51}\times10^{-2}$&$25.8^{+11.3}_{-7.3}$&$2.93\pm0.19$&$4.63^{+2.50}_{-1.50}\times10^{-1}$&$1625^{+925}_{-625}$&294.1/242\\
7&F&$0.89\pm0.09$&$2.35^{+0.68}_{-0.58}$&$0.90^{+0.19}_{-0.20}$&$1.50^{+0.34}_{-0.29}\times10^{-1}$&$<4.80\times10^{-1}$&-&-&-&730.2/746\\
8&B&-&-&-&-&$1.96^{+1.47}_{-1.45}$&$1.77\pm0.07$&$13.1^{+0.93}_{-0.85}$&-&844.0/828\\
9&B&-&-&-&-&$1.84^{+0.46}_{-0.45}$&$1.70\pm0.06$&$1.43^{+0.09}_{-0.08}$&-&1834.5/2211\\
10&C&-&-&-&-&$21.5^{+4.8}_{-4.5}$&$1.45^{+0.16}_{-0.15}$&$5.36^{+0.86}_{-0.79}$&$169^{+149}_{-80}$&743.6/807\\
11&D&-&-&$1.58\pm0.19$&$1.41^{+0.53}_{-0.38}\times10^{-1}$&$15.6^{+2.7}_{-2.4}$&$1.58\pm0.19$&$1.30^{+0.19}_{-0.17}$&-&1153.7/1371\\
12&C&-&-&-&-&$12.2^{+2.4}_{-2.7}$&$1.42^{+0.09}_{-0.14}$&$9.39^{+0.94}_{-1.22}$&$<235$&777.5/832\\
13&A&-&-&$1.75\pm0.03$&$1.20\pm0.06$&$<4.99\times10^{-2}$&-&-&-&2239.2/2492\\
14&D&-&-&$1.86\pm0.23$&$2.17^{+1.06}_{-0.72}\times10^{-2}$&$16.6^{+2.3}_{-2.2}$&$1.86\pm0.23$&$1.16^{+0.18}_{-0.16}$&-&1049.0/1168\\
15&D&-&-&$1.75^{+0.40}_{-0.38}$&$2.52^{+2.52}_{-1.23}\times10^{-2}$&$16.2^{+2.6}_{-2.3}$&$1.75^{+0.40}_{-0.38}$&$5.33^{+1.20}_{-0.61}$&-&60.7/54*\\
16&B&-&-&-&-&$4.08^{+1.51}_{-1.25}$&$1.22^{+0.60}_{-0.50}$&$1.09^{+0.17}_{-0.13}$&-&20.1/21*\\
17&E&-&-&$1.09\pm0.71$&$2.44^{+5.62}_{-1.70}\times10^{-2}$&$7.79^{+3.47}_{-2.93}$&$1.09\pm0.71$&$5.14^{+1.50}_{-1.13}\times10^{-1}$&$1457^{+1180}_{-680}$&118.5/170\\
18&B&-&-&-&-&$1.66^{+1.64}_{-1.51}\times10^{-1}$&$2.43^{+0.50}_{-0.46}$&$3.42^{+1.41}_{-1.03}\times10^{-2}$&-&136.7/112\\
19&D&-&-&$1.38\pm0.09$&$1.72^{+0.37}_{-0.30}\times10^{-1}$&$1.48\pm0.16$&$1.38\pm0.09$&$8.79^{+0.60}_{-0.48}\times10^{-1}$&-&207.8/209*\\
20&G&$0.84\pm0.02$&$7.67^{+0.63}_{-0.61}$&-&-&$7.97^{+4.66}_{-4.43}\times10^{-2}$&$2.17\pm0.16$&$1.07^{+0.13}_{-0.12}\times10^{-1}$&-&1665.0/1868\\
\end{tabular}
\end{table}
\clearpage
 \begin{table}
 \centering
 \begin{tabular}[H]{ccccccccccc}
 \multicolumn{11}{l}{\textbf{Table \ref{result}}\,(Continued)}\\
     \hline
   &&\multicolumn{2}{c}{apec}&\multicolumn{2}{r}{None absorbed pegpwrlw}&zphabs&\multicolumn{2}{r}{absorbed pegpwrlw}&\multicolumn{1}{c}{zgauss}&\\
   \hline
   No.&Model&{\it kT}&Norm&$\Gamma_u$&Norm1&Intrinsic~$N\rm{_H}$&$\Gamma_a$&Norm2&EW&${\it C}\rm{-stat}/\rm{d.o.f.}$\\
   &&keV&$10^{-5}$&&$10^{-12}\,\rm{erg\,cm^{-2}\,s^{-1}}$&$10^{22}\rm{cm^{-2}}$&&$10^{-12}\,\rm{erg\,cm^{-2}\,s^{-1}}$&eV&\\
   (1)&(2)&(3)&(4)&(5)&(6)&(7)&(8)&(9)&(10)&(11)\\
    \hline
21&C&-&-&-&-&$38.6^{+13.8}_{-13.5}$&$1.31^{+0.32}_{-0.37}$&$2.72^{+1.25}_{-0.89}$&$469^{+245}_{-174}$&647.3/747\\
22&A&-&-&$1.09\pm0.41$&$8.90^{+5.77}_{-3.63}\times10^{-2}$&$<8.42\times10^{-1}$&-&-&-&33.4/43\\
23&A&-&-&$0.68^{+0.27}_{-0.28}$&$2.11^{+0.97}_{-0.71}\times10^{-1}$&$<6.01\times10^{-1}$&-&-&-&142.20/136\\
24&A&-&-&$1.67\pm0.05$&$4.31^{+0.29}_{-0.28}\times10^{-1}$&$<4.18\times10^{-2}$&-&-&-&432.1/375\\
25&D&-&-&$1.93^{+0.27}_{-0.26}$&$3.99^{+1.89}_{-1.31}\times10^{-2}$&$90.1^{+25.2}_{-22.0}$&$1.93^{+0.27}_{-0.26}$&$1.12^{+0.72}_{-0.45}$&-&391.1/428\\
26&A&-&-&$2.04^{+0.28}_{-0.26}$&$2.56^{+0.83}_{-0.67}\times10^{-2}$&$<1.27\times10^{-1}$&-&-&-&97.0/138\\
27&F&$0.83\pm0.02$&$7.90^{+0.75}_{-0.71}$&$1.47\pm0.11$&$2.00^{+0.22}_{-0.20}\times10^{-1}$&$<6.98\times10^{-3}$&-&-&-&1352.6/1513\\
28&B&-&-&-&-&$1.80^{+0.59}_{-0.0.57}\times10^{-1}$&$1.85\pm0.09$&$2.71^{+0.27}_{-0.25}\times10^{-1}$&-&1371.4/1576\\
29&F&$0.74\pm0.06$&$2.45^{+0.45}_{-0.43}$&$2.14^{+0.14}_{-0.15}$&$5.33^{+0.85}_{-0.76}\times10^{-2}$&$<1.20\times10^{-1}$&-&-&-&1131.1/1256\\
30&F&$0.91^{+0.07}_{-0.06}$&$2.01^{+0.41}_{-0.38}$&$2.33^{+0.03}_{-0.04}$&$1.48\pm0.06\times10^{-1}$&$<1.75\times10^{-2}$&-&-&-&2039.9/2263\\
31&A&-&-&$1.67\pm0.07$&$1.30^{+0.18}_{-0.16}\times10^{-1}$&$<8.76\times10^{-2}$&-&-&-&785.0/959\\
32&A&-&-&$1.68^{+0.37}_{-0.35}$&$5.39^{+3.02}_{-2.07}\times10^{-2}$&$<3.29\times10^{-1}$&-&-&-&52.2/50\\
33&A&-&-&$1.750\pm0.005$&$10.1\pm0.1$&$<3.92\times10^{-4}$&-&-&-&5300.99/5294\\
34&G*&$0.88^{+0.02}_{-0.01}$&$48.4\pm0.2$&-&-&$4.87^{+2.45}_{-1.97}$&$2.27^{+0.56}_{-0.50}$&$2.55^{+0.65}_{-0.39}\times10^{-1}$&-&2132.5/2203\\
35&A&-&-&$1.82\pm0.01$&$7.12\pm0.10$&$<1.64\times10^{-3}$&-&-&-&4663.7/4666\\
36&A&-&-&$1.78\pm0.01$&$8.38\pm0.10$&$<1.24\times10^{-1}$&-&-&-&5047.38/4919\\
37&A&-&-&$1.53\pm0.10$&$4.39^{+0.55}_{-0.50}$&$<1.35\times10^{-1}$&-&-&-&280.3/333\\
38&B&-&-&-&-&$9.37^{+0.48}_{-0.32}\times10^{-1}$&$1.56\pm0.04$&$6.60\pm0.14$&-&4468.1/4663\\
39&H&$0.75\pm0.01$&$30.1^{+1.2}_{-1.1}$&$1.70^{+0.18}_{-0.19}$&$1.59^{+0.05}_{-0.04}\times10^{-1}$&$8.73^{+1.36}_{-1.31}$&$1.70^{+0.18}_{-0.19}$&$9.65^{+1.27}_{-1.23}\times10^{-1}$&-&2412.1/2710\\
40&A&-&-&$2.26\pm0.04$&$1.53\pm0.09$&$<2.45\times10^{-2}$&-&-&-&2072.8/2190\\
\end{tabular}
\end{table}
\clearpage
 \begin{table}
 \centering
 \begin{tabular}[H]{ccccccccccc}
 \multicolumn{11}{l}{\textbf{Table \ref{result}}\,(Continued)}\\
     \hline
   &&\multicolumn{2}{c}{apec}&\multicolumn{2}{r}{None absorbed pegpwrlw}&zphabs&\multicolumn{2}{r}{absorbed pegpwrlw}&\multicolumn{1}{c}{zgauss}&\\
   \hline
   No.&Model&{\it kT}&Norm&$\Gamma_u$&Norm1&Intrinsic~$N\rm{_H}$&$\Gamma_a$&Norm2&EW&${\it C}\rm{-stat}/\rm{d.o.f.}$\\
   &&keV&$10^{-5}$&&$10^{-12}\,\rm{erg\,cm^{-2}\,s^{-1}}$&$10^{22}\rm{cm^{-2}}$&&$10^{-12}\,\rm{erg\,cm^{-2}\,s^{-1}}$&eV&\\
   (1)&(2)&(3)&(4)&(5)&(6)&(7)&(8)&(9)&(10)&(11)\\
    \hline
41&A&-&-&$2.12^{+0.29}_{-0.28}$&$2.54^{+1.08}_{-0.78}\times10^{-1}$&$<3.22\times10^{-1}$&-&-&-&51.2/81\\
42&A&-&-&$1.49\pm0.04$&$2.19\pm0.11$&$<1.34\times10^{-2}$&-&-&-&393.7/438\\
43&A&-&-&$1.54\pm0.19$&$9.05^{+2.38}_{-1.95}\times10^{-2}$&$<1.53\times10^{-1}$&-&-&-&125.3/141\\
44&A&-&-&$1.67\pm0.01$&$5.43^{+0.11}_{-0.10}$&$<4.07\times10^{-3}$&-&-&-&4230.7/4445\\
45&A&-&-&$2.49\pm0.02$&$2.85^{+0.08}_{-0.07}$&$<1.35\times10^{-2}$&-&-&-&2547.4/2809\\
46&A&-&-&$1.47\pm0.18$&$1.14^{+0.27}_{-0.23}$&$<4.52\times10^{-1}$&-&-&-&120.8/151\\
47&A&-&-&$1.37\pm0.15$&$2.48^{+0.43}_{-0.37}$&$<1.24\times10^{-1}$&-&-&-&208.6/206\\
48&A&-&-&$1.74\pm0.10$&$7.66^{+1.17}_{-1.06}\times10^{-1}$&$<4.06\times10^{-2}$&-&-&-&1759.5/1964\\
49&A&-&-&$1.70^{+0.11}_{-0.10}$&$2.22^{+0.18}_{-0.17}$&$<6.33$&-&-&-&501.4/539\\
50&F&$0.79\pm0.03$&$4.08^{+0.46}_{-0.44}$&$2.08\pm0.06$&$1.36^{+0.11}_{-0.10}\times10^{-1}$&$<1.17\times10^{-2}$&-&-&-&1698.3/2002\\
51&A&-&-&$2.26^{+0.17}_{-0.16}$&$3.82^{+1.07}_{-0.90}\times10^{-2}$&$<8.36\times10^{-2}$&-&-&-&383.7/496\\
52&A&-&-&$1.57\pm0.11$&$6.18^{+0.91}_{-0.81}$&$<4.65\times10^{-2}$&-&-&-&287.3/280\\
53&G&$0.70\pm0.01$&$23.8\pm1.1$&-&-&$6.57^{+1.62}_{-1.65}\times10^{-1}$&$1.82\pm0.12$&$7.50^{+0.40}_{-0.39}\times10^{-1}$&-&2598.5/2747\\
54&A&-&-&$1.88\pm0.11$&$1.35^{+0.20}_{-0.18}$&$<1.12\times10^{-1}$&-&-&-&209.0/218\\
55&A&-&-&$1.74\pm0.02$&$1.97\pm0.08$&$<4.46\times10^{-3}$&-&-&-&2664.7/2928\\
56&A&-&-&$1.70\pm0.04$&$3.82^{+0.25}_{-0.24}\times10^{-1}$&$<1.94\times10^{-2}$&-&-&-&2875.2/3121\\
57&A&-&-&$2.19\pm0.12$&$4.41^{+0.75}_{-0.65}\times10^{-1}$&$<1.14\times10^{-1}$&-&-&-&162.6/179\\
58&A&-&-&$2.11\pm0.02$&$4.79\pm0.13$&$<4.69\times10^{-3}$&-&-&-&2833.1/3144\\
59&A&-&-&$1.99^{+0.40}_{-0.39}$&$9.14^{+4.55}_{-3.17}\times10^{-1}$&$<4.66\times10^{-1}$&-&-&-&39.7/63\\
60&F&$0.62^{+0.04}_{-0.05}$&$7.03^{+0.87}_{-0.84}$&$2.18\pm0.07$&$2.21^{+0.20}_{-0.18}\times10^{-1}$&$<1.29\times10^{-2}$&-&-&-&1447.6/1643\\
\end{tabular}
\end{table}
\clearpage
 \begin{table}
 \centering
 \begin{tabular}[H]{ccccccccccc}
 \multicolumn{11}{l}{\textbf{Table \ref{result}}\,(Continued)}\\
     \hline
   &&\multicolumn{2}{c}{apec}&\multicolumn{2}{r}{None absorbed pegpwrlw}&zphabs&\multicolumn{2}{r}{absorbed pegpwrlw}&\multicolumn{1}{c}{zgauss}&\\
   \hline
   No.&Model&{\it kT}&Norm&$\Gamma_u$&Norm1&Intrinsic~$N\rm{_H}$&$\Gamma_a$&Norm2&EW&${\it C}\rm{-stat}/\rm{d.o.f.}$\\
   &&keV&$10^{-5}$&&$10^{-12}\,\rm{erg\,cm^{-2}\,s^{-1}}$&$10^{22}\rm{cm^{-2}}$&&$10^{-12}\,\rm{erg\,cm^{-2}\,s^{-1}}$&eV&\\
   (1)&(2)&(3)&(4)&(5)&(6)&(7)&(8)&(9)&(10)&(11)\\
    \hline
61&A&-&-&$2.06^{+0.19}_{-0.18}$&$5.97^{+1.45}_{-1.20}\times10^{-1}$&$<1.47\times10^{-1}$&-&-&-&125.6/141\\
62&B&-&-&-&-&$3.34^{+0.85}_{-0.84}\times10^{-2}$&$2.53\pm0.03$&$1.44^{+0.04}_{-0.03}$&-&30146.8/3278\\
63&A&-&-&$2.61^{+0.83}_{-0.76}$&$1.18^{+1.66}_{-0.72}\times10^{-1}$&$<2.70\times10^{-1}$&-&-&-&23.0/27\\
64&A&-&-&$1.72\pm0.01$&$10.7\pm0.1$&$<1.69\times10^{-3}$&-&-&-&4732.4/4842\\
65&A&-&-&$3.49^{+0.63}_{-0.59}$&$1.17^{+1.08}_{-0.58}\times10^{-2}$&$<1.51\times10^{-1}$&-&-&-&40.6/39\\
66&A&-&-&$1.75\pm0.03$&$3.03^{+0.14}_{-0.13}$&$<3.78\times10^{-2}$&-&-&-&2493.7/2722\\
67&D&-&-&$1.58^{+0.44}_{-0.45}$&$2.57^{+2.87}_{-1.30}\times10^{-2}$&$4.90^{+1.42}_{-1.26}$&$1.58^{+0.44}_{-0.45}$&$2.56^{+0.51}_{-0.46}\times10^{-1}$&-&176.2/237\\
68&A&-&-&$2.44^{+0.38}_{-0.36}$&$1.08^{+0.67}_{-0.43}$&$<6.45\times10^{-1}$&-&-&-&53.0/65\\
\hline
\multicolumn{11}{l}{(1) Object number in Table \ref{sample}. (2) Spectral model; A: {\tt phabs*pegpwrlw}~B: {\tt phabs*zphabs*pegpwrlw}~C: {\tt phabs*(zphabs*pegpwrlw+zgauss)}}\\
\multicolumn{11}{l}{D: {\tt phabs*(pegpwrlw+zphabs*pegpwrlw)}~E: {\tt phabs*(pegpwrlw+zphabs*pegpwrlw+zgauss)}~F: {\tt phabs*(apec+pegpwrlw)}}\\
\multicolumn{11}{l}{G: {\tt phabs*(apec+zphabs*pegpwrlw)}~H: {\tt phabs*(apec+pegpwrlw+zphabs*pegpwrlw)}, {\it XMM-Newton} and {\it NuSTAR} have multiple datasets}\\
\multicolumn{11}{l}{(e.g., {\it XMM-Newton} has three data sets: mos1, mos2, and pn), and thus the models are multiplied by const. We also multiplied the absorbing}\\
\multicolumn{11}{l}{edge model (zedge) for the No. 34 source, whose model is marked with *. (3) Temperature of the plasma. (4) Normalization of the apec model.}\\
\multicolumn{11}{l}{(5) Photon index of unabsorbed power-law component. (6) Flux of unabsorbed power-law component in 2--10\,keV. (7) Intrinsic hydrogen column}\\
\multicolumn{11}{l}{density. (8) Photon index of absorbed power-law component. (9) Flux of absorbed power-law component in 2--10\,keV.}\\
\multicolumn{11}{l}{(10) Equivalent width of the Iron emission line. (11) {\it C}\rm-statistics and degree of freedom. In the case of the analysis using Sherpa (noted by *),}\\
\multicolumn{11}{l}{the binning min was set to 1 and used $\chi^2$.}\\
\end{tabular}
\end{table}
\end{landscape}

\begin{table}[h]
    \caption{X-ray luminosity and Eddington luminosity ratio}
    \label{Edington}
    \centering
    \begin{tabular}{ccccccc}
    \hline
    No.&$L_{2-10}$&$L_{2-10}/L{\rm_{Edd}}$&&No.&$L_{2-10}$&$L_{2-10}/L{\rm_{Edd}}$\\
    &$\rm{erg\,s^{-1}}$&&&&$\rm{erg\,s^{-1}}$&\\
        \cline{1-3} \cline{5-7}
    (1)&(2)&(3)&&(1)&(2)&(3)\\
    \cline{1-3}\cline{5-7}
1&$1.17^{+0.10}_{-0.07}\times10^{44}$&$5.61^{+0.48}_{-0.36}\times10^{-4}$&&35&$1.30\pm0.02\times10^{43}$&$7.45^{+0.11}_{-0.10}\times10^{-4}$\\
2&$5.21^{+0.65}_{-0.59}\times10^{44}$&$3.94^{+0.49}_{-0.45}\times10^{-3}$&&36&$6.40\pm0.07\times10^{43}$&$4.46\pm0.05\times10^{-4}$\\
3&$2.18^{+0.44}_{-0.34}\times10^{44}$&$1.90^{+0.38}_{-0.29}\times10^{-3}$&&37&$9.19^{+1.16}_{-1.05}\times10^{45}$&$2.62^{+0.33}_{-0.30}\times10^{-2}$\\
4&$4.61^{+0.29}_{-0.25}\times10^{43}$&$5.38^{+0.34}_{-0.29}\times10^{-4}$&&38&$2.47\pm0.05\times10^{42}$&$1.98\pm0.04\times10^{-4}$\\
5&$2.49^{+0.40}_{-0.35}\times10^{44}$&$1.48^{+0.24}_{-0.21}\times10^{-3}$&&39&$1.32^{+0.15}_{-0.14}\times10^{41}$&$2.27^{+0.26}_{-0.25}\times10^{-5}$\\
6&$1.36^{+0.70}_{-0.42}\times10^{43}$&$7.81^{+4.05}_{-2.42}\times10^{-5}$&&40&$1.62^{+0.10}_{-0.09}\times10^{42}$&$2.30^{+0.14}_{-0.13}\times10^{-5}$\\
7&$2.14^{+0.48}_{-0.41}\times10^{43}$&$8.69^{+1.95}_{-1.66}\times10^{-5}$&&41&$8.89^{+3.78}_{-2.74}\times10^{44}$&$2.23^{+0.95}_{-0.69}\times10^{-2}$\\
8&$2.42^{+0.17}_{-0.16}\times10^{44}$&$2.29^{+0.16}_{-0.15}\times10^{-3}$&&42&$8.93^{+0.46}_{-0.44}\times10^{45}$&$1.66\pm0.08\times10^{-2}$\\
9&$1.08^{+0.07}_{-0.06}\times10^{45}$&-&&43&$2.86^{+0.75}_{-0.62}\times10^{42}$&$1.08^{+0.28}_{-0.23}\times10^{-5}$\\
10&$4.47^{+0.72}_{-0.66}\times10^{43}$&$3.86^{+0.62}_{-0.57}\times10^{-4}$&&44&$2.32\pm0.04\times10^{45}$&-\\
11&$8.16^{+1.01}_{-0.86}\times10^{44}$&$3.31^{+0.41}_{-0.35}\times10^{-3}$&&45&$2.02\pm0.05\times10^{43}$&$1.25\pm0.03\times10^{-4}$\\
12&$7.00^{+0.70}_{-0.91}\times10^{43}$&$4.71^{+0.47}_{-0.61}\times10^{-4}$&&46&$2.22^{+0.53}_{-0.44}\times10^{45}$&$1.76^{+0.42}_{-0.35}$\\
13&$9.54^{+0.48}_{-0.46}\times10^{44}$&-&&47&$6.58^{+1.14}_{-0.99}\times10^{45}$&$8.29^{+1.43}_{-1.25}\times10^{-2}$\\
14&$4.73^{+0.74}_{-0.65}\times10^{44}$&$2.57^{+0.40}_{-0.35}\times10^{-3}$&&48&$7.74^{+1.18}_{-1.07}\times10^{43}$&$4.55^{+0.69}_{-0.63}\times10^{-4}$\\
15&$3.12^{+0.70}_{-0.35}\times10^{44}$&$2.40^{+0.54}_{-0.27}\times10^{-3}$&&49&$1.18\pm0.09\times10^{44}$&$2.95^{+0.23}_{-0.22}\times10^{-3}$\\
16&$1.13^{+0.17}_{-0.13}\times10^{43}$&$2.50^{+0.38}_{-0.30}\times10^{-4}$&&50&$5.30^{+0.42}_{-0.40}\times10^{41}$&$1.76^{+0.14}_{-0.13}\times10^{-5}$\\
17&$3.57^{+0.99}_{-0.75}\times10^{42}$&$3.01^{+0.84}_{-0.63}\times10^{-5}$&&51&$2.13^{+0.60}_{-0.50}\times10^{41}$&$8.83^{+2.47}_{-2.08}\times10^{-6}$\\
18&$2.76^{+1.14}_{-0.83}\times10^{41}$&$6.54^{+2.70}_{-1.97}\times10^{-7}$&&52&$2.71^{+0.40}_{-0.35}\times10^{44}$&$2.07^{+0.30}_{-0.27}\times10^{-3}$\\
19&$1.38^{+0.09}_{-0.07}\times10^{43}$&$1.84^{+0.12}_{-0.10}\times10^{-4}$&&53&$4.58^{+0.25}_{-0.24}\times10^{41}$&$3.83^{+0.21}_{-0.20}\times10^{-5}$\\
20&$7.00^{+0.82}_{-0.75}\times10^{40}$&$6.77^{+0.79}_{-0.73}\times10^{-7}$&&54&$2.69^{+0.40}_{-0.36}\times10^{44}$&$1.71^{+0.26}_{-0.23}\times10^{-3}$\\
21&$5.37^{+2.48}_{-1.77}\times10^{43}$&$1.04^{+0.48}_{-0.34}\times10^{-3}$&&55&$1.29\pm0.05\times10^{43}$&-\\
22&$4.24^{+2.75}_{-1.73}\times10^{41}$&$4.41^{+2.86}_{-1.80}\times10^{-6}$&&56&$1.78^{+0.12}_{-0.11}\times10^{41}$&$1.46^{+0.10}_{-0.09}\times10^{-5}$\\
23&$1.87^{+0.85}_{-0.62}\times10^{44}$&$4.30^{+1.97}_{-1.43}\times10^{-4}$&&57&$4.07^{+0.69}_{-0.60}\times10^{41}$&$2.01^{+0.34}_{-0.30}\times10^{-5}$\\
24&$3.11^{+0.21}_{-0.20}\times10^{43}$&$1.45^{+0.10}_{-0.09}\times10^{-4}$&&58&$1.27^{+0.04}_{-0.03}\times10^{44}$&$1.63^{+0.05}_{-0.04}\times10^{-3}$\\
25&$4.46^{+2.79}_{-1.70}\times10^{44}$&$1.08^{+0.67}_{-0.31}\times10^{-3}$&&59&$1.08^{+0.54}_{-0.37}\times10^{42}$&$8.09^{+4.03}_{-2.80}\times10^{-5}$\\
26&$7.82^{+2.52}_{-2.04}\times10^{41}$&$2.98^{+0.96}_{-0.78}\times10^{-6}$&&60&$3.32^{+0.30}_{-0.28}\times10^{41}$&$1.83^{+0.16}_{-0.15}\times10^{-5}$\\
27&$2.79^{+0.30}_{-0.27}\times10^{41}$&$1.80^{+0.19}_{-0.18}\times10^{-6}$&&61&$5.04^{+1.23}_{-1.01}\times10^{41}$&$2.74^{+0.67}_{-0.55}\times10^{-5}$\\
28&$2.21^{+0.22}_{-0.20}\times10^{44}$&$5.26^{+0.52}_{-0.48}\times10^{-4}$&&62&$1.16\pm0.03\times10^{42}$&$7.11^{+0.18}_{-0.17}\times10^{-5}$\\
\end{tabular}
\end{table}
\clearpage
\begin{table}[t]
\centering
\begin{tabular}{ccccccc}
 \multicolumn{7}{l}{\textbf{Table \ref{Edington}}\,(Continued)}\\
    \hline
    No.&$L_{2-10}$&$L_{2-10}/L{\rm_{Edd}}$&&No.&$L_{2-10}$&$L_{2-10}/L{\rm_{Edd}}$\\
    &$\rm{erg\,s^{-1}}$&&&&$\rm{erg\,s^{-1}}$&\\
        \cline{1-3} \cline{5-7}
    (1)&(2)&(3)&&(1)&(2)&(3)\\
    \cline{1-3}\cline{5-7}
29&$8.20^{+1.30}_{-1.16}\times10^{40}$&$3.59^{+0.57}_{-0.51}\times10^{-6}$&&63&$2.55^{+3.57}_{-1.54}\times10^{41}$&$7.60^{+10.65}_{-4.61}\times10^{-6}$\\
30&$9.15^{+0.39}_{-0.38}\times10^{41}$&$3.20^{+0.14}_{-0.13}\times10^{-6}$&&64&$5.75\pm0.07\times10^{43}$&$3.87\pm0.05\times10^{-3}$\\
31&$1.20^{+0.16}_{-0.15}\times10^{43}$&$8.77^{+1.19}_{-1.07}\times10^{-5}$&&65&$1.35^{+1.24}_{-0.66}\times10^{40}$&$7.58^{+6.99}_{-3.74}\times10^{-7}$\\
32&$5.79^{+3.24}_{-2.22}\times10^{42}$&$1.14^{+0.64}_{-0.44}\times10^{-4}$&&66&$1.44\pm0.06\times10^{44}$&$1.53\pm0.07\times10^{-3}$\\
33&$2.86\pm0.02\times10^{43}$&$1.88\pm0.02\times10^{-3}$&&67&$7.30^{+1.31}_{-1.19}\times10^{40}$&$9.83^{+1.77}_{-1.60}\times10^{-6}$\\
34&$1.97^{+0.50}_{-0.31}\times10^{40}$&$2.88^{+0.73}_{-0.45}\times10^{-6}$&&68&$4.91^{+3.06}_{-1.95}\times10^{44}$&$1.40^{+0.87}_{-0.56}\times10^{-6}$\\
\hline
\multicolumn{7}{l}{(1) Object number in Table \ref{sample}. (2) Absorption-corrected luminosity at 2--10\,keV.}\\
\multicolumn{7}{l}{(3) Eddington luminosity ratio.}
\end{tabular}
\end{table}

\begin{table}[H]
 \caption{Fraction of absorbed RGs ($N\rm{_H}>10^{22}cm^{-2}$)}
 \label{absratio}
 \centering
  \begin{tabular}{ccccc}
  
   \hline
   & FR-I & FR-II & CSS & total\\
   \hline
   GeV-quiet & 0.18 & 0.70 & 1 & 0.53\\
   GeV-loud & 0.16 & 0 & 0 & 0.08\\
   \hline 
   total & 0.17 & 0.44 & 0.17 & 0.29\\
   \hline
  \end{tabular}
\end{table}

\begin{figure}[t]
  \begin{minipage}[b]{0.5\hsize}
    \centering
    {3C\,032}
    \includegraphics[height=7.5cm,width=9cm]{modelD.pdf}
    \label{fermia}
    \end{minipage} 
  \begin{minipage}[b]{0.5\hsize}
    \centering
    {3C\,403}
    \includegraphics[height=7.5cm,width=9cm]{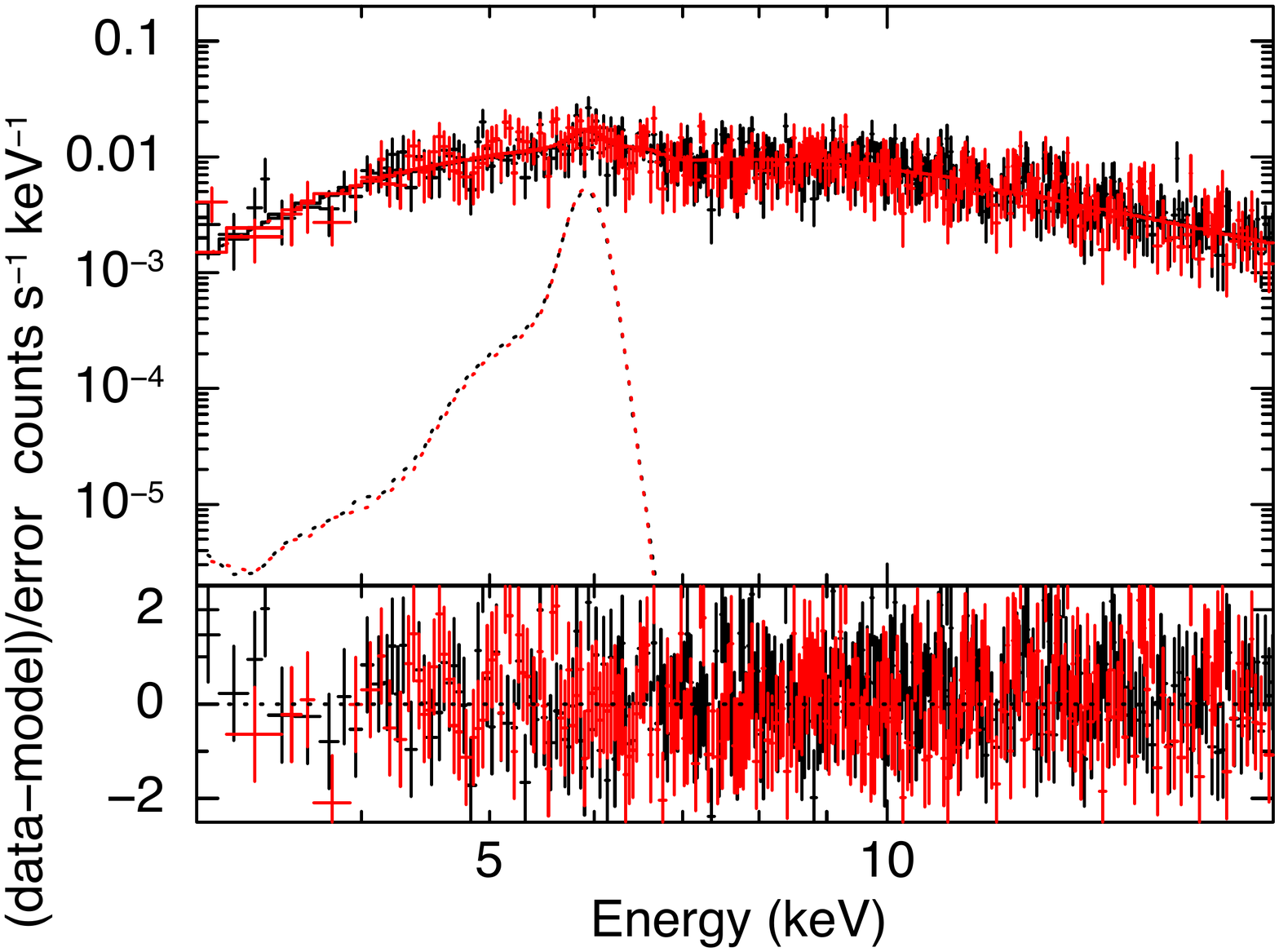}
    \label{fermib}
  \end{minipage} \\
  
        \begin{minipage}[b]{0.5\hsize}
        \centering
        {3C\,327.1}
        \includegraphics[height=7.5cm,width=9cm]{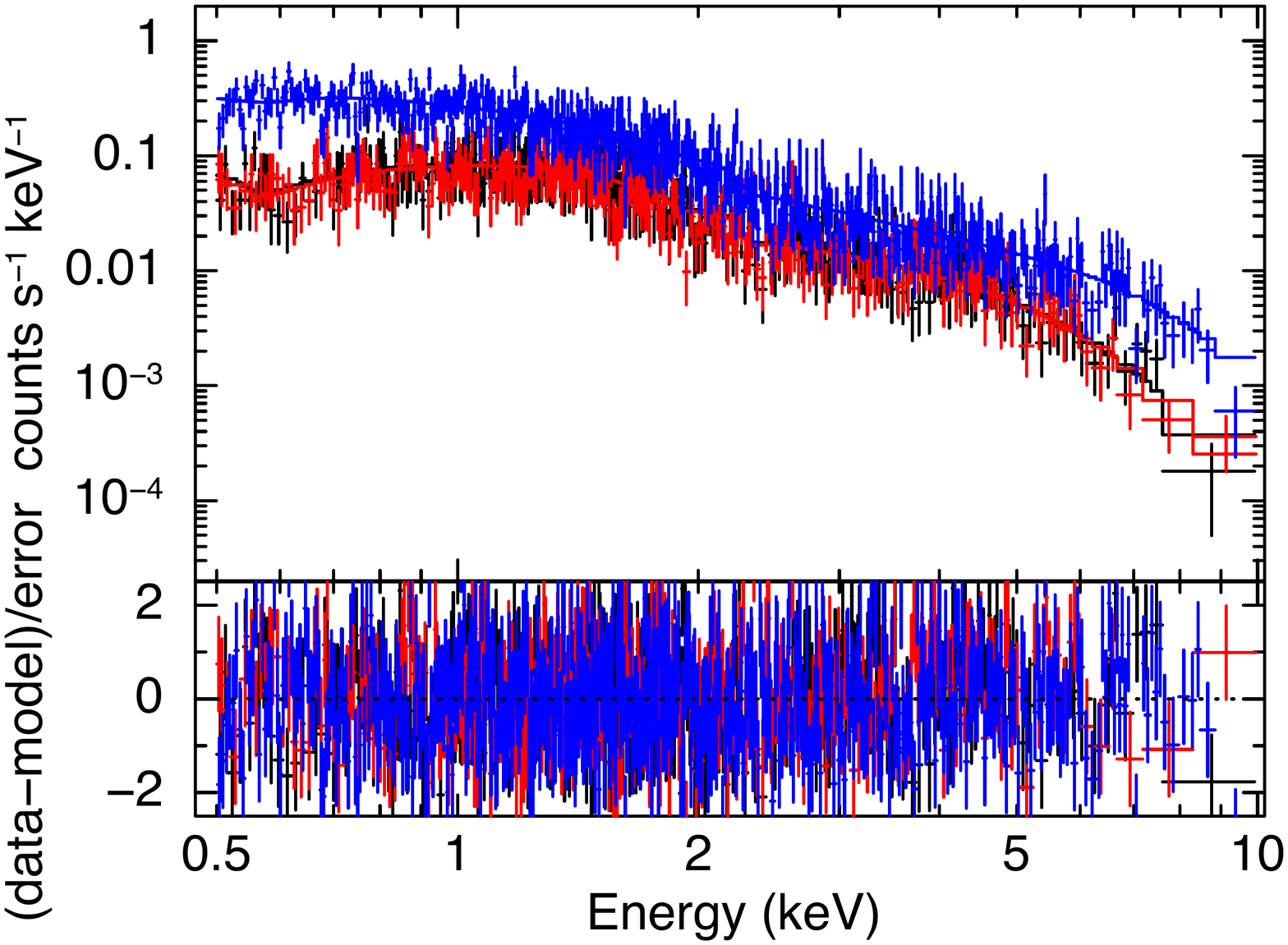}
        \vspace*{-0.5cm}
        \label{fermic}
      \end{minipage}
      \begin{minipage}[b]{0.5\hsize}
        \centering
        {PKS\,0039-44}
        \includegraphics[height=7.5cm,width=9cm]{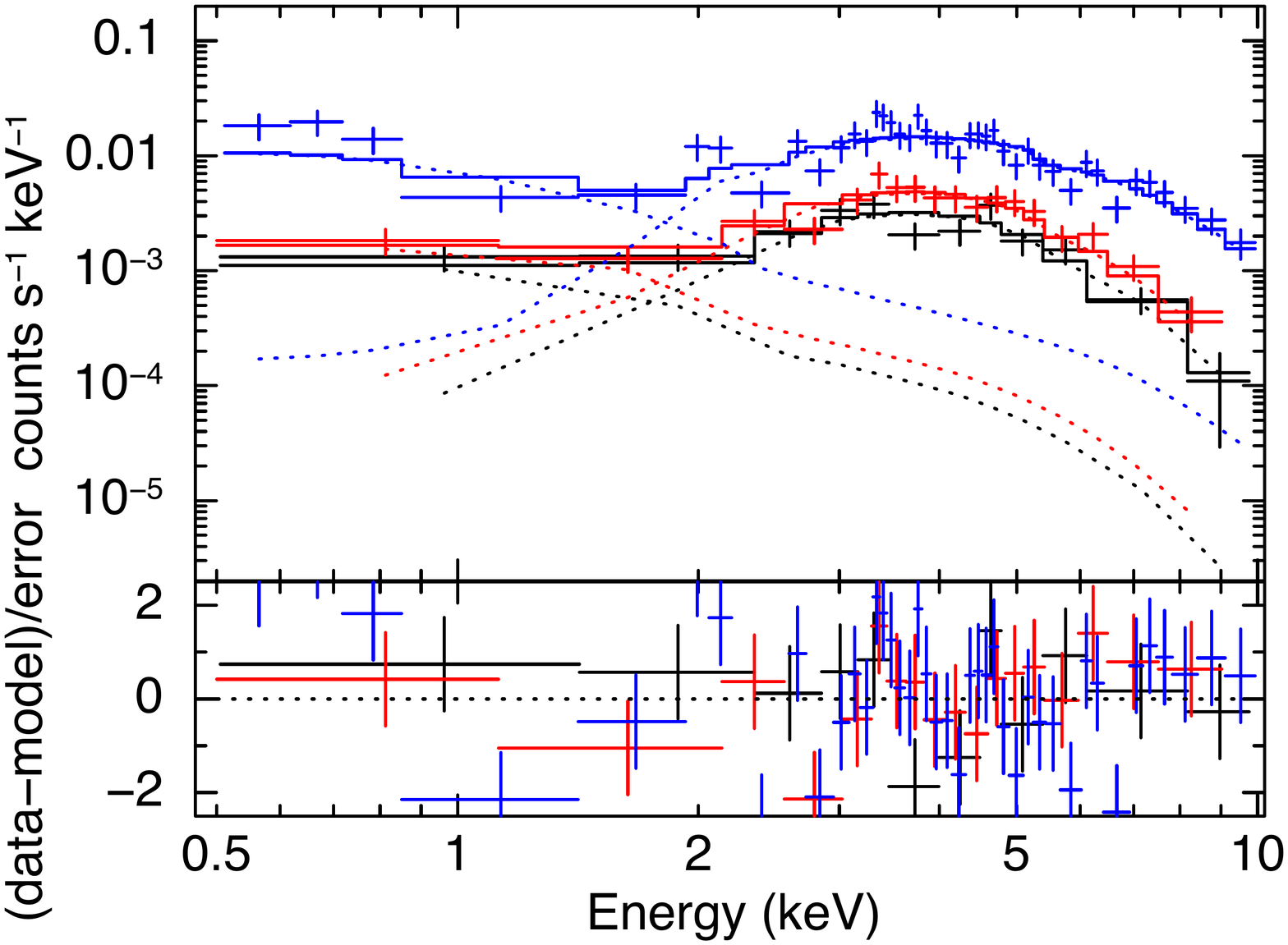}
        \vspace*{-0.5cm}
        \label{fermid}
      \end{minipage}\\
     \caption{Example of X-ray spectra of GeV-quiet RGs. Details are the same as Figure \ref{mode1a-h}.}
     \label{nonefermispec}
\end{figure}

\begin{figure}[t]
     \begin{minipage}[b]{0.5\hsize}
        \centering
        {IC\,1531}
        \includegraphics[height=7.5cm,width=9cm]{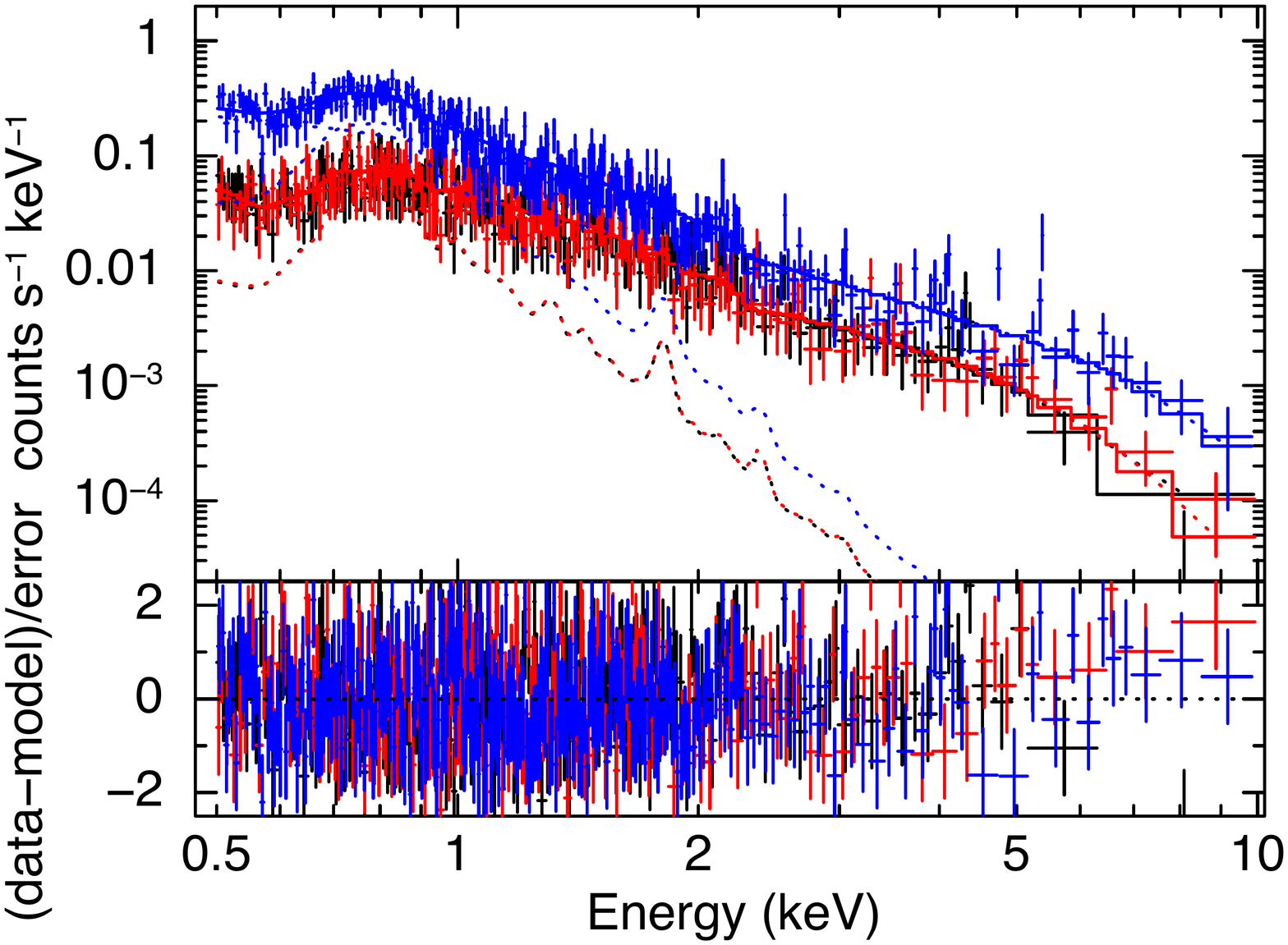}
        \vspace*{-0.5cm}
        \label{fermie}
      \end{minipage} 
      \begin{minipage}[b]{0.5\hsize}
        \centering
        {PKS\,0625-35}
        \includegraphics[height=7.5cm,width=9cm]{modelA.pdf}
        \vspace*{-0.5cm}
        \label{fermif}
      \end{minipage} \\
  
      \begin{minipage}[b]{0.5\hsize}
        \centering
        {Pictor\,A}
        \includegraphics[height=7.5cm,width=9cm]{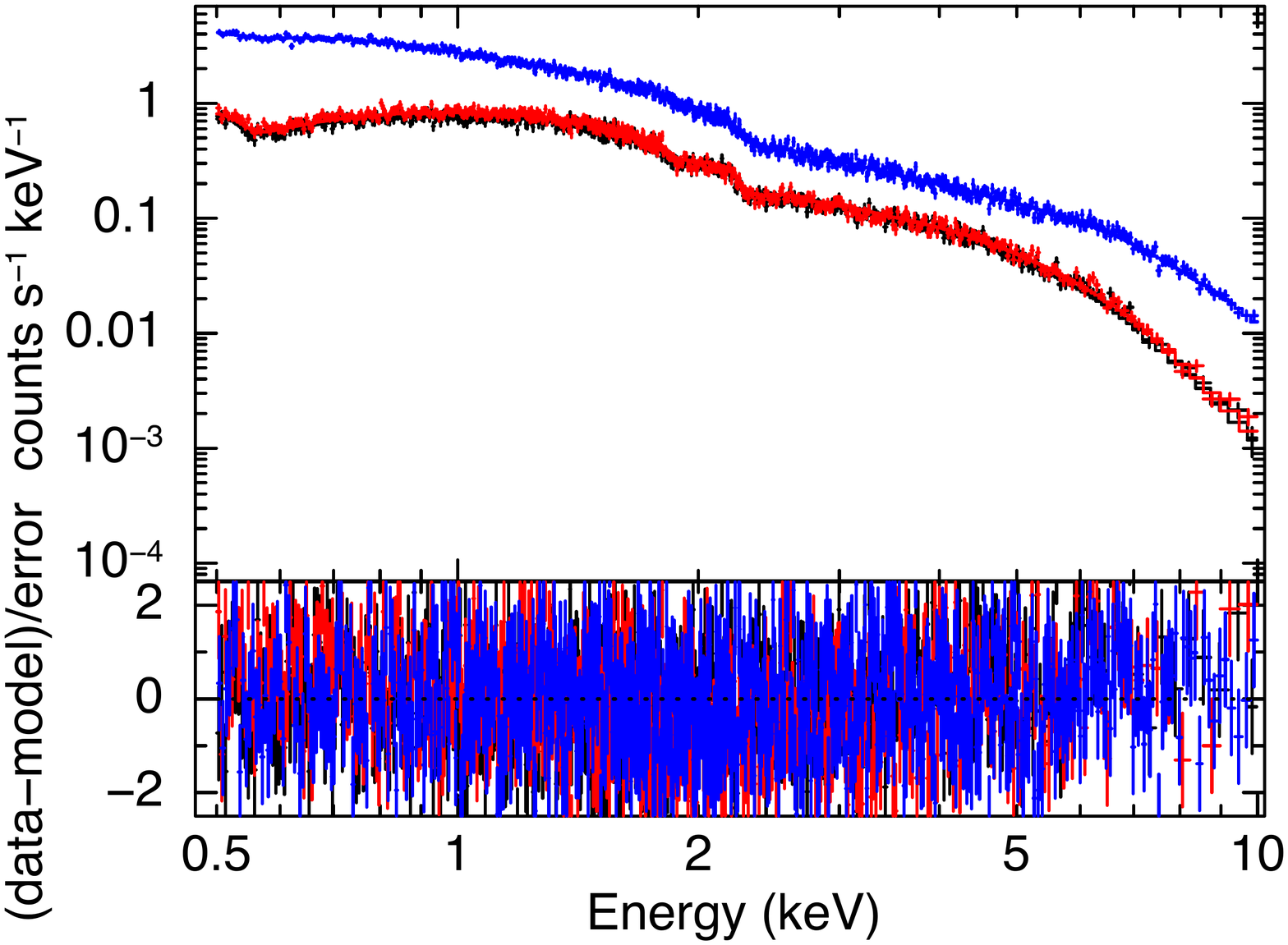}
        \vspace*{-0.5cm}
        \label{fermig}
      \end{minipage} 
      \begin{minipage}[b]{0.5\hsize}
        \centering
        {PKS\,1514+00}
        \includegraphics[height=7.5cm,width=9cm]{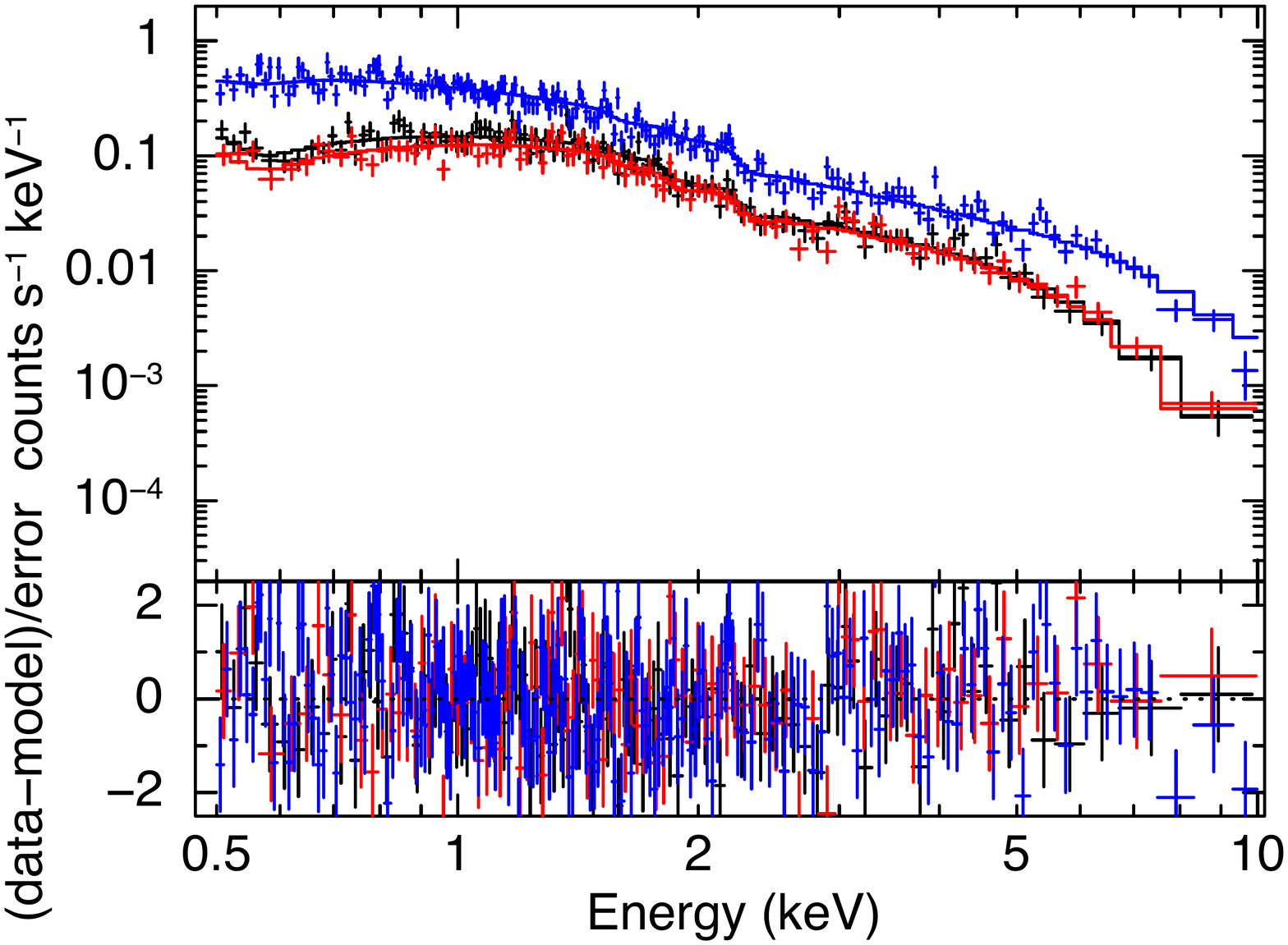}
        \vspace*{-0.5cm}
        \label{fermih}
      \end{minipage} \\

     \caption{Example of X-ray spectra of GeV-loud RGs. Details are the same as Figure \ref{mode1a-h}.}
     \label{fermispec}
\end{figure}
\clearpage

\end{document}